\documentclass[aps,pra,preprint,groupedaddress,twocolumn,10pt]{revtex4-1}
\usepackage{graphicx}
\usepackage{epsfig}
\usepackage{epstopdf}
\usepackage{subfigure}
\usepackage{appendix}
\usepackage{amsmath,xfrac,amssymb}
\begin{document}
\title{Light deflection by light: Effect of incidence angle and inhomogeneity}
\author{Pardeep Kumar}

\email[]{pradeep.kumar@iitrpr.ac.in}
\author{Shubhrangshu Dasgupta}

\affiliation{Department of Physics, Indian Institute of Technology Ropar, Rupnagar, Punjab 140001, India}

\date{\today}

\begin{abstract}
We study the angular deflection of the circularly polarized components of a linearly polarized probe field in a weakly birefringent atomic system in tripod configuration.  A spatially inhomogeneous control field incident obliquely onto an atomic vapor cell facilitates a large angular divergence between circular components. We show that the angular resolution can be dynamically controlled by optimally choosing the angle of incidence and the transverse profile of the control beam. For instance, by employing a Laguerre-Gaussian profile of the control field, one can impart a large angular divergence to the circular components close to the entry face of the atomic vapor cell. We further demonstrate how such a medium causes the focusing and refocusing of the probe field, thereby acting as  a lens with multiple foci. The absorption in the medium remains negligible at resonance due to electromagnetically induced transparency (EIT). 
\end{abstract}


\maketitle

\section{Introduction}
Many optical phenomena like refraction and dispersion involve a change in the trajectory of the incident light as a  consequence of the spatial variation of the refractive index of the medium. In the past decades, optical beam deflection has achieved a considerable attention. The deflection of light beam can be achieved by  mechanical interaction \cite{cindrich1967}, thermal gradient \cite{jackson1981}, acousto-optical interaction \cite{dixon1967}, electro-optic effect \cite{lee1968} and all-optical methods \cite{agarwal1990}. Optical methods have enjoyed much attention due to their high speed, efficiency and fast nonlinear response time.

Manoeuvering light by another light through their interaction with the medium has thus created a new avenue of research. Much interest has been given to the deflection of light beam in a homogeneous medium subjected to external fields with spatially inhomogeneous intensity distributions. The spatial modulation of the refractive index of the medium induced by a suitable inhomogeneous control field leads to several effects such as dynamic light deflection \cite{moseley1995,moseley1996}, waveguiding \cite{kapoor2000}, and antiwaveguiding \cite{arbiv2001}. A significant variation of the refractive index at resonance can give rise to lage deflection.  Recently, such a deflection in an atomic medium, exhibiting electromagnetically induced transparency (EIT), is observed in presence of a \textit{magnetic field} with small gradient transverse to the propagation direction \cite{weis1992,holzner1997,karpa2006}. In another related experiment \cite{scully2010}, it is found that the light ray can also be deflected when an \textit{optical field} with inhomogeneous transverse profile drives a cell with Rb atomic vapors with  $\Lambda$-type energy-level configuration. The angle of deviation obtained by means of optical field is found to be much larger (by several order) than that reported using inhomogeneous magnetic field \cite{karpa2006}. An adequate explanation for the observed phenomenon of light deflection can be provided in the framework of semiclassical theory \cite{sun2006,zhou2007}. Further, the beam deflection in EIT medium can also be explained quantum mechanically in terms of dark state \textit{polariton} possessing an effective magnetic moment \cite{karpa2006,zhou2008,zhang2009}. Besides this, the deflection of light  in terms of \textit{vector optical solitons} (nonlinear polariton) in a double EIT system is proposed in \cite{hang2012}. Further, in comparison to EIT medium, active Raman gain medium (ARG) is shown to induce larger deflection of light beams \cite{zhu2013} and antiwaveguiding \cite{onkar2015}.

In order to measure the small deflections of beams, many sophisticated interferometric setups have been proposed \cite{hoston2008,dixon2009}. Some of these techniques are inspired by quantum \textit{weak measurement} \cite{aharonov1988} to resolve Angstrom-scale optical beam deflection in space as well as in time domain \cite{brunner2003,pardeep2015}. In this paper, instead of making sensitive detection of small deflection, we rather propose a way to increase the angular separation between the circular components of the probe field by using an inhomogeneous control field in an atomic vapor system with tripod configuration. We present a theoretical analysis (in the framework of \textit{semiclassical} theory) to demonstrate light deflection where the control beam is incident obliquely to the entrance face of the vapor cell. This oblique incidence of the control beam causes the mixing of the excitations on all the three optical transitions in the tripod system, leading to an extra flexibility to produce large angular divergence among $\sigma_{\pm}$ components of the probe field by changing the angle of incidence. It should be borne in mind that a medium of chiral molecules also creates the birefringence and hence the angular divergence between the circular components of the linearly polarized light. In \cite{ghosh2006}, it has been demonstrated that such an effect can occur at the interface of achiral and chiral media and can be used for the detection of optical activity \cite{condon1937} with a \textit{miniaturized} sample volume. Here, we mimic such a situation in \textit{atomic vapors} by exploiting large angular divergence {\it close to the entry face\/} of the vapor cell.

Note that an inhomogeneous control field can create a birefringence inside a medium \cite{guo2008}. By suitably choosing the transverse profile of the control field, one can enhance this birefringence to impart larger angular divergence to the orthogonal components. We show that by choosing a Laguerre-Gaussian  profile \cite{allen1992} of the control field, angular divergence can be made larger, compared to that as obtained using a Gaussian profile. 
This also results in the focusing and refocusing effects of $\sigma_{\pm}$ components of the probe field at various axial positions. We detect a few focal points for the probe field depending on the profile and the incidence angle of the control beam. Thereby, a medium with suitable length behaves as a lens \cite{zhang2009} with multiple foci. It is to be emphasized that the large angular divergence of the circular components of the probe field is accompanied by zero absorption at resonance, thanks to EIT \cite{kumar2013}. This suggests a way how to obtain sufficiently large deflection, even with an EIT medium, instead of an ARG medium.

The organization of the paper is as follows. In Sec. II, the theoretical model is introduced along with the eikonal approximation for the probe deflection. The influence of the angle of incidence and the profile of control field on the deflection and the transmission of circular components of the probe field is described in Sec. III. In this Section, the main results of the paper are presented. Sec. IV summarizes this paper.

\section{Physical Model}
\begin{figure}[h!]
\begin{center}
\begin{tabular}{l}
\subfigure[]{\includegraphics[scale=0.37]{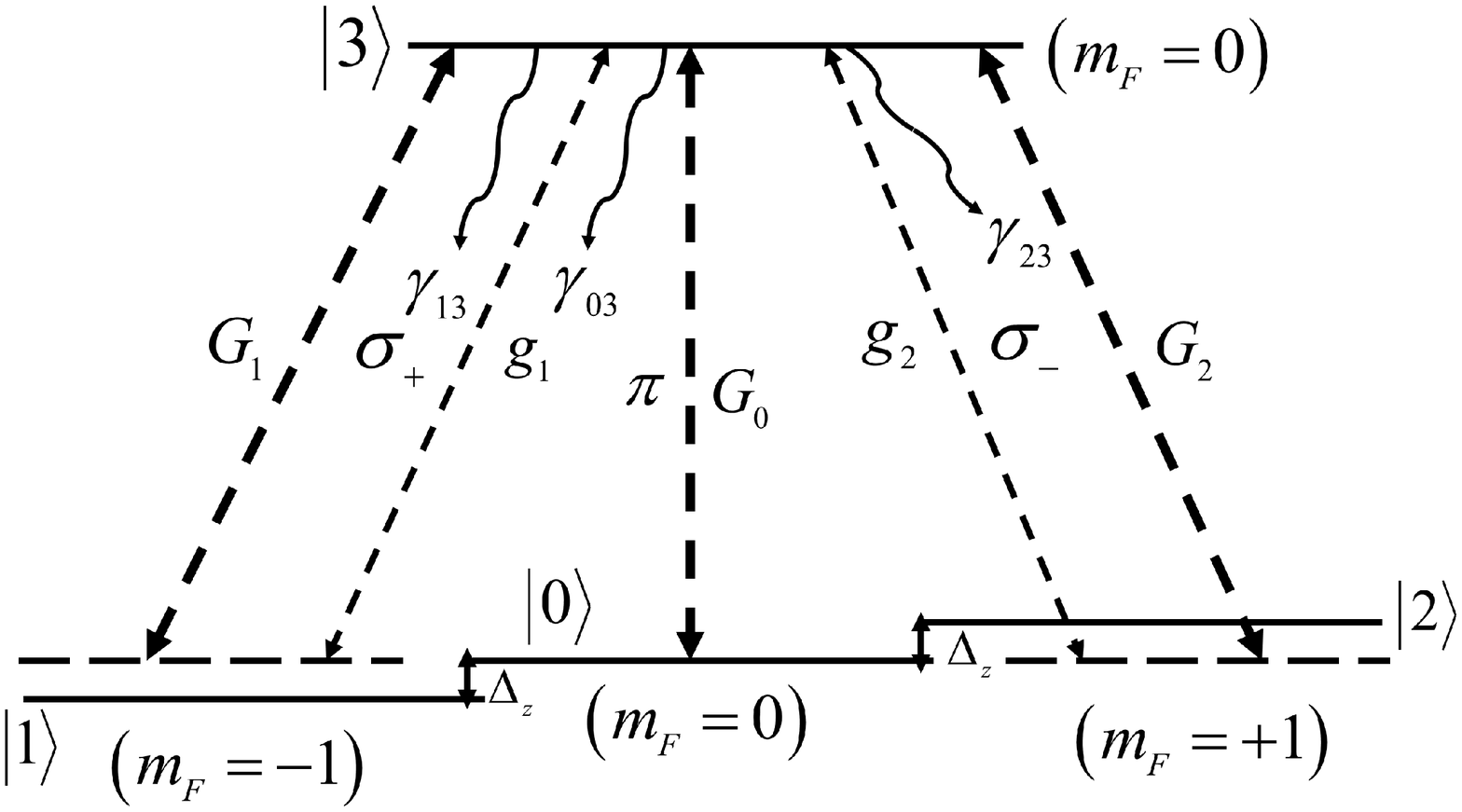}}\\ \subfigure[]{\includegraphics[scale=0.23]{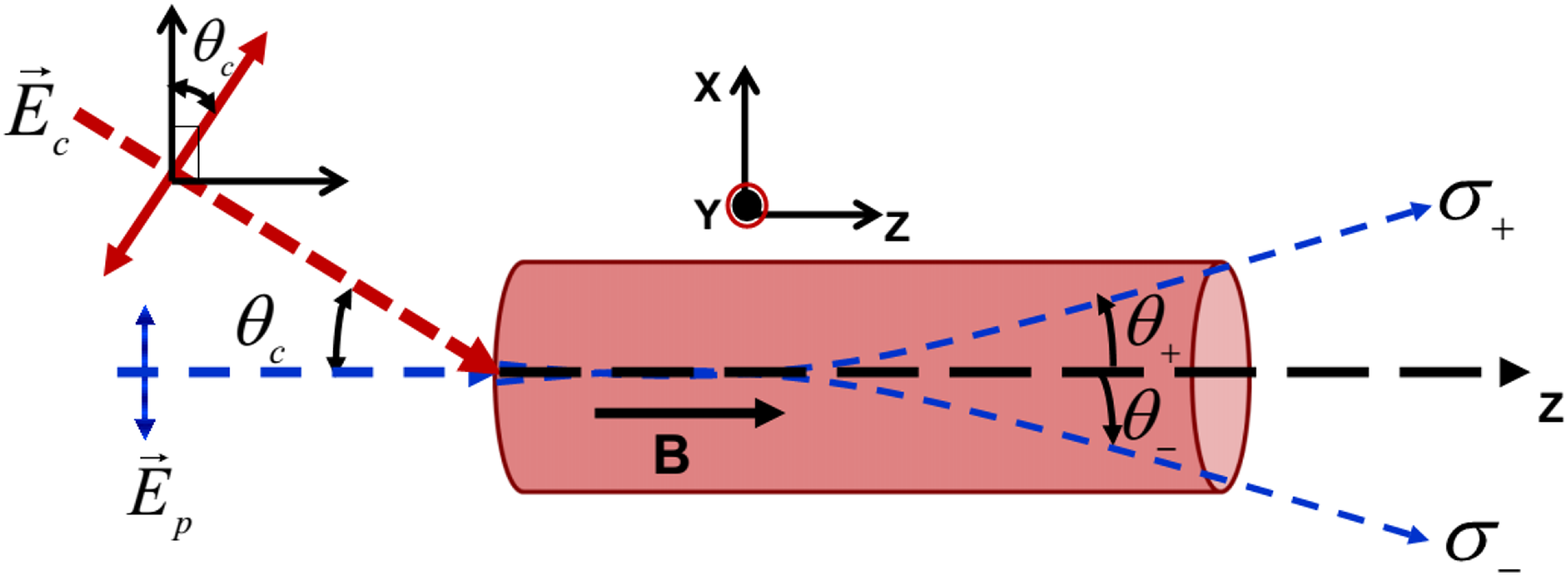}}
\end{tabular}
\caption{(Color online) (a) Energy-level configuration. The $\sigma_{\pm}$ components of the probe field drive the transitions $|1\rangle\leftrightarrow|3\rangle$ and $|2\rangle\leftrightarrow|3\rangle$, respectively. A control field incident at an angle $\theta_{c}$ to the entry face excites all the transitions of the tripod system. The degeneracy of the ground levels is removed by applying a feeble axial magnetic field. $\gamma_{j3} \left(j=0,1,2\right)$ are the spontaneous decay rates from $|3\rangle$ to $|j\rangle.$ (b) Schematic for the deflection of the circular components of the probe field.}
\label{fig1}
\end{center}
\end{figure}

We consider a generic four-level tripod atomic system as shown in Fig. \ref{fig1}(a). The relevant energy levels can be found in many systems such as in $^{39}$K \cite{steck}, $^{23}$Na \cite{dasgupta2002}, $^{7}$Li \cite{arnold2001}, Pr:YSO \cite{ham2000}. Here, we used the transition $^{2}S_{1/2}\rightarrow {}^{2}P_{1/2}$ at 769.9 nm of $^{39}$K vapors. The ground levels are $|1\rangle={}^{2}S_{1/2}\left(F=2, m_{F}=-1\right)$, $|0\rangle={}^{2}S_{1/2}\left(F=2, m_{F}=0\right)$, $|2\rangle={}^{2}S_{1/2}\left(F=2, m_{F}=+1\right)$ and the upper state is $|3\rangle={}^{2}P_{1/2}\left(F=1, m_{F}=0\right)$. The upper level $|3\rangle \left(m_{F}=0\right)$ is coupled to the ground levels $|1\rangle \left(m_{F}=+1\right)$ and $|2\rangle \left(m_{F}=-1\right)$ by a $\hat{x}$-polarized weak probe field $\vec{E}_{p}=\hat{x}\varepsilon_{p}e^{-i\left(\omega_{p}t-k_{p}z\right)}+c.c.$ propagating along $z$-direction. Here,  $\varepsilon_{p}$ is the slowly varying envelope, $\omega_{p}$ is the angular frequency and $k_{p}$ is the propagation constant of the probe field. The orthogonal components of the probe with $\sigma_{+}$ and $\sigma_{-}$ polarizations couple to $|1\rangle\leftrightarrow|3\rangle$ and $|2\rangle\leftrightarrow|3\rangle$ transitions, respectively. The Rabi frequencies of the corresponding transitions are defined as 2$g_{1}=$2$\left(\frac{\vec{d}_{31}\cdot\hat{x}\varepsilon_{p}}{\hbar}\right)$ and 2$g_{2}=$2$\left(\frac{\vec{d}_{32}\cdot\hat{x}\varepsilon_{p}}{\hbar}\right)$, where $\vec{d}_{ij}$ represents the transition electric dipole moment matrix element between the levels $|i\rangle$ and $|j\rangle$.  The degeneracy of the ground level has been removed by applying a dc magnetic field of strength $B$ along the quantization $z$-axis. The Zeeman splitting among the respective levels is $\Delta_{z}=\mu_{B}Bmg_{F}/h$, where $\mu_{B}$ is Bohr magneton and $g_{F}$ is the hyperfine Land\'{e} $g$-factor. A control field $\vec{E}_{c}=\varepsilon_{c}\left(\cos\theta_{c}\hat{x}+\sin\theta_{c}\hat{z}\right)e^{ik_{c}\left(-x\sin\theta_{c}+z\cos\theta_{c}\right)-i\omega_{c} t}+c.c.$  is incident obliquely at an angle $\theta_{c}$ to the medium with polarization lying in the plane of incidence, as shown in Fig. \ref{fig1} (b). Here $\varepsilon_{c}$, $k_{c}$ and $\omega_{c}$ are the slowly varying envelope, propagation constant and the frequency of the control field. This field couples to $|j\rangle\leftrightarrow|3\rangle \left(j=0,1,2\right)$ transitions. The corresponding Rabi frequencies of the control field are given by 2$G_{i}$=2$\left[\varepsilon_{c}\left(\cos\theta_{c}\hat{x}+\sin\theta_{c}\hat{z}\right)\right]\cdot \vec{d}_{ji}/\hbar$ ($i=$0,1,2 and $j=$3). Using the circular polarization vector $\hat{\epsilon}=\left(\hat{x}\pm i\hat{y}\right)/\sqrt{2}$, these Rabi frequencies can be simplified as 
\begin{align}\label{eq1}
G_{1}=\frac{|\vec{d}|\varepsilon_{c}}{\hbar\sqrt{2}}\cos\theta_{c},~G_{2}=\frac{|\vec{d}|\varepsilon_{c}}{\hbar\sqrt{2}}\cos\theta_{c},~G_{0}=\frac{|\vec{d}_{30}|\varepsilon_{c}}{\hbar}\sin\theta_{c}\;,
\end{align}
where,  we have chosen $|\vec{d}_{31}|=|\vec{d}_{32}|=|\vec{d}|$. 

The Hamiltonian for the above system in dipole approximation can be written as

\begin{equation}
\begin{array}{lll}
\hat{H}&=&\sum\limits_{j=1}^{3}\hbar\omega_{j0}|j\rangle\langle j|-\sum\limits_{k=0}^{2}\left(G_{k}e^{-i\omega_{c}t}|3\rangle\langle k|+\mbox{H.c.}\right)\\&-&\sum\limits_{k=1}^{2}\left(g_{k}e^{-i\omega_{p}t}|3\rangle\langle k|+\mbox{H.c.}\right)\;.\\
\end{array}
\label{eq2}
\end{equation}
Here zero of energy is defined at the level $|0\rangle$ and $\hbar\omega_{\alpha\beta}$ is the energy difference between the levels $|\alpha\rangle$ and $|\beta\rangle$. We describe the dynamical evolution of the system by the density matrix equations in Markovian limit, as given in the Appendix.

\subsection{Susceptibility of the atomic medium}
In the steady state, we can obtain the approximate solution for the linear susceptibility of the medium for the probe field. Here, we are interested in the atomic coherences $\tilde{\rho}_{31}^{(+1)}$ and $\tilde{\rho}_{32}^{(-1)}$  for the orthogonal components of the probe field which can be obtained by solving Eq. (\ref{eqa3}). Thus, the susceptibility of the atomic medium for the orthogonal components of the probe field is given by
\begin{align}\label{eq3}
\chi_{+} &= \frac{3Nc^{3}}{2\omega_{p}^{3}}\tilde{\rho}_{31}^{\prime(+1)}, & \chi_{-} &= \frac{3Nc^{3}}{2\omega_{p}^{3}}\tilde{\rho}_{32}^{\prime(-1)}\;.
\end{align}
where, $N$ is the number density of the atomic medium and $c$ is the speed of light in vacuum.
\subsection{Probe deflection}
In general, the spatial structure of the applied fields results in the position dependence of the medium susceptibility [Eq. (\ref{eq3})]. The trajectory of a light ray propagating through an inhomogeneous medium can be found by solving  an eikonal equation \cite{born1999}
\begin{align}\label{eq4}
\left(\nabla\psi\right)\cdot\left(\nabla\psi\right)=n^{2}\;,
\end{align}
where, the eikonal $\psi$ represents the phase of the electromagnetic wave and $n=1+2\pi\mbox{Re}\left[\chi\right]$ describes the refractive index of the medium. Now, by defining $\nabla\psi=n\frac{d\vec{R}}{ds}$, we obtain following differential equation: 
\begin{align}\label{eq5}
\frac{d}{ds}\left(n\frac{d\vec{R}}{ds}\right)=\nabla n \;.
\end{align}
Here, $\vec{R}=X\left(z\right)\hat{e}_{x}+Y\left(z\right)\hat{e}_{y}+z\hat{e}_{z}$ is a point on the light ray and $ds=\sqrt{dx^{2}+dy^{2}+dz^{2}}$. In component form, Eq. (\ref{eq5}) yields
\begin{align}\label{eq6}
\frac{d}{ds}\left(n\frac{dX}{ds}\right)=\frac{\partial n}{\partial x}, \frac{d}{ds}\left(n\frac{dY}{ds}\right)=\frac{\partial n}{\partial y}, \frac{d}{ds}\left(n\frac{dz}{ds}\right)=\frac{\partial n}{\partial z}\;.
\end{align}
In  paraxial limit, $ds\approx dz$ for small deflections and the first two equations in Eq. (\ref{eq6}) reduce to an ordinary differential equation describing the ray trajectories, as follows:
\begin{align}\label{eq7}
\frac{d^{2}X}{dz^{2}}=\frac{\partial n}{\partial x}~~\mbox{and}~~\frac{d^{2}Y}{dz^{2}}=\frac{\partial n}{\partial y}\;.
\end{align}
Using Eq. (\ref{eq7}) the trajectory of the ray and the deflection angle can be estimated. Let us assume that the atomic vapor cell can be divided into many smaller cells such that the external inhomogeneous field appears to be homogeneous for each smaller cell. Thus, the angle of deflection for the probe field \cite{scully2010} can be obtained from Eq. (\ref{eq7}) by solving
\begin{align}\label{eq8}
\frac{d\tan\theta_{x}}{dz}=\frac{\partial n}{\partial x}~~\mbox{and}~~\frac{d\tan\theta_{y}}{dz}=\frac{\partial n}{\partial y}\;,
\end{align}
where, $\tan\theta_{x}\left(\tan\theta_{y}\right)$ represents the slope  and $\theta_{x}\left(\theta_{y}\right)$ is the angle of deflection of the light rays in the xz-plane (yz-plane). For smaller angle of deflection, $\tan\theta_{x}\approx\theta_{x}\left(\tan\theta_{y}\approx\theta_{y}\right)$ and Eq. (\ref{eq8}) yields
\begin{align}\label{eq9}
\theta_{x}=\int_{0}^{L}dz\frac{\partial n}{\partial x}~~ \mbox{and}~~\theta_{y}=\int_{0}^{L}dz\frac{\partial n}{\partial y}\;,
\end{align}
where, $L$ is the length of the medium in the direction of propagation. In this paper, we consider that the transverse profile of the control field is confined to the $y=0$ plane. Due to the anisotropy induced by the magnetic field and the inhomogeneous control field \cite{guo2008}, the refractive index $n_{\pm}$ and the corresponding angle of deflection $\theta_{\pm}$ of the $\sigma_{\pm}$ components will be different, giving rise to an angular divergence $\phi=\theta_{+}-\theta_{-}$.
\subsection{Transmission of circular components}
Further, the imaginary part of the susceptibilities $\chi_{\pm}$ [Eq. (\ref{eq3})] determines the transmission of the right and left circularly polarized components of the probe field. These susceptibilities depend upon the transverse profile of the control field, and therefore, the transmission of the circular components can be written as
\begin{align}\label{eq10}
T_{\pm}\left(x,y,L\right)=\mbox{exp}\left\{-k_{p}\int_{0}^{L}\mbox{Im}\left[\chi_{\pm}\left(x,y,z;\omega_{p}\right)\right]dz\right\}\;,
\end{align}
which under the paraxial approximation reduces to
\begin{align}\label{eq11}
T_{\pm}\left(L\right)=\mbox{exp}\left\{-k_{p}\int_{0}^{L}\mbox{Im}\left[\chi_{\pm}\left(z;\omega_{p}\right)\right]dz\right\}\;.
\end{align}
\section{Results}
\subsection{Beam profile of the control field}
In order to produce the spatially dependent refractive index for the probe field, we choose the following transverse profile of the control field 
\begin{align}\label{eq12}
\varepsilon_{c}(x,y,z)&=\varepsilon_{0}\frac{w_{0}}{w_{z}}\left(\frac{\sqrt{2}r}{w_{z}}\right)^{m}e^{-\frac{r^{2}}{w_{z}^{2}}}\exp\left[-\frac{ikr^{2}}{2R_{z}}+im\theta\right]\;\nonumber\\
&\times\exp\left[-i\left(m+1\right)\tan^{-1}\left(\frac{Z}{z_{R}}\right)\right]\;,
\end{align}
where, $\varepsilon_{0}$ is the initial peak amplitude, $w_{z}=w_{0}\sqrt{1+\left(Z/z_{R}\right)^{2}}$ is the beam width with $w_{0}$ as the beam waist at $Z=0$ and $z_{R}=\pi w_{0}^{2}/\lambda$ is the Rayleigh length. Here, $r=\sqrt{X^{2}+y^{2}}$ is the radial distance from the axis of the beam, $\theta=\tan^{-1}\left(\frac{y}{X}\right)$ and $R_{z} =Z\left(1+\left(\frac{z_{R}}{Z}\right)^{2}\right)$. As the control field is incident at an angle $\theta_{c}$ to the vapor cell, we can write $X=x\cos\theta_{c}-z\sin\theta_{c}$ and $Z=x\sin\theta_{c}+z\cos\theta_{c}$ as the coordinates in the plane of incidence. For $m=0$, the above profile becomes Gaussian, and takes the following simplified form 
\begin{align}\label{eq13}
\varepsilon_{c}(x,y,z)=\varepsilon_{0}^{\prime}\mbox{exp}\left(-\frac{X^{2}+y^{2}}{\sigma^{2}}\right)\;.
\end{align}
where, $\varepsilon_{0}^{\prime}$ defines the maximum amplitude of the Gaussian beam and $\sigma$ refers to its transverse width. Note that the full width at half maxima (FWHM) of the above profile is $2\sqrt{2\ln2}\sigma$.

On the other hand, $m\neq 0$ corresponds to a Laguerre-Gaussian (LG)$_{m}$ profile with azimuthal index $m$.  Such a spatial structure of the control field produces inhomogeneity for the susceptibility of the probe field along the transverse direction and results in its deflection.  
\subsection{Dependence of angular divergence on angle of incidence}
\begin{figure}[ht!]
\begin{center}
\begin{tabular}{cc}
\includegraphics[scale=0.23]{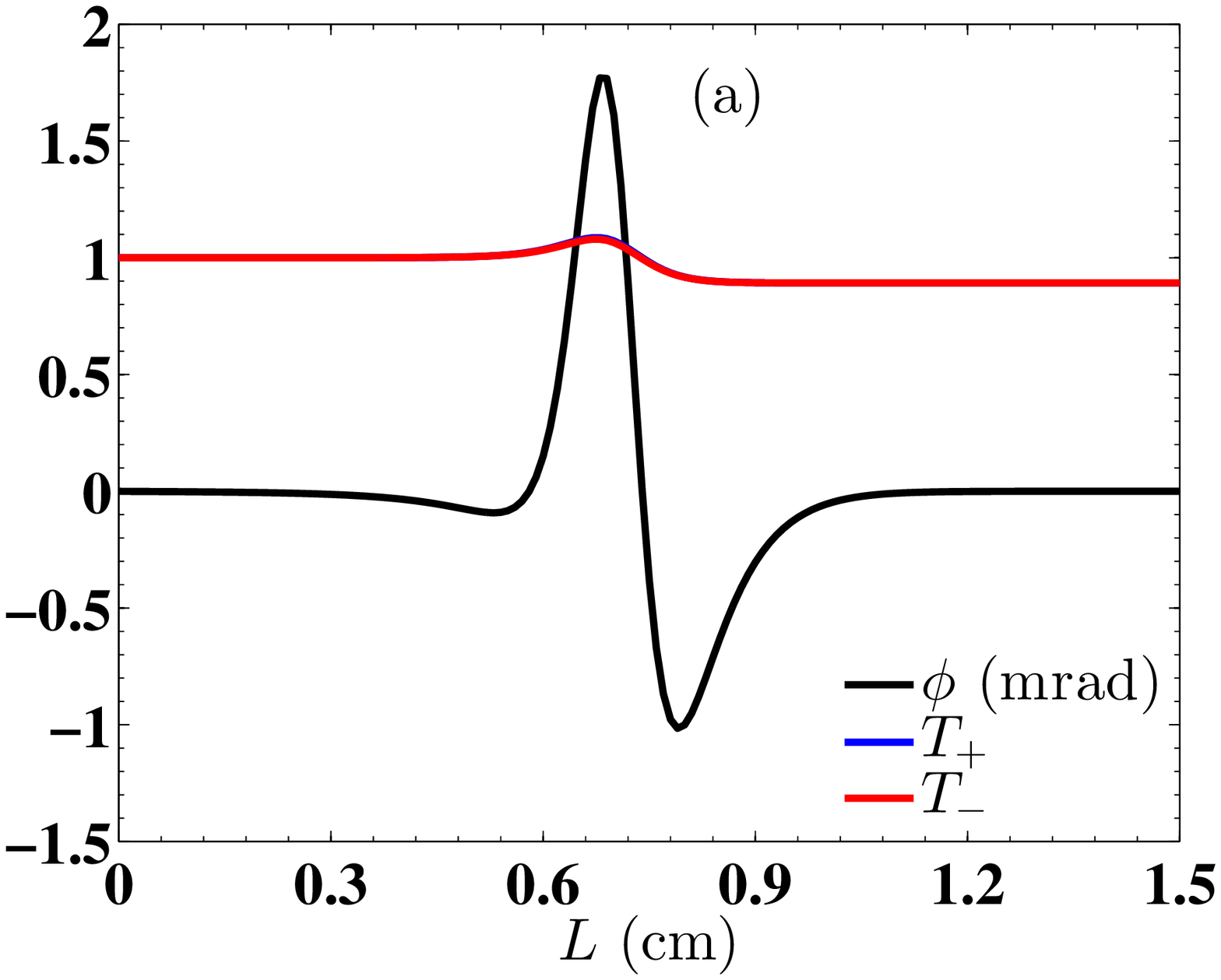}& \includegraphics[scale=0.23]{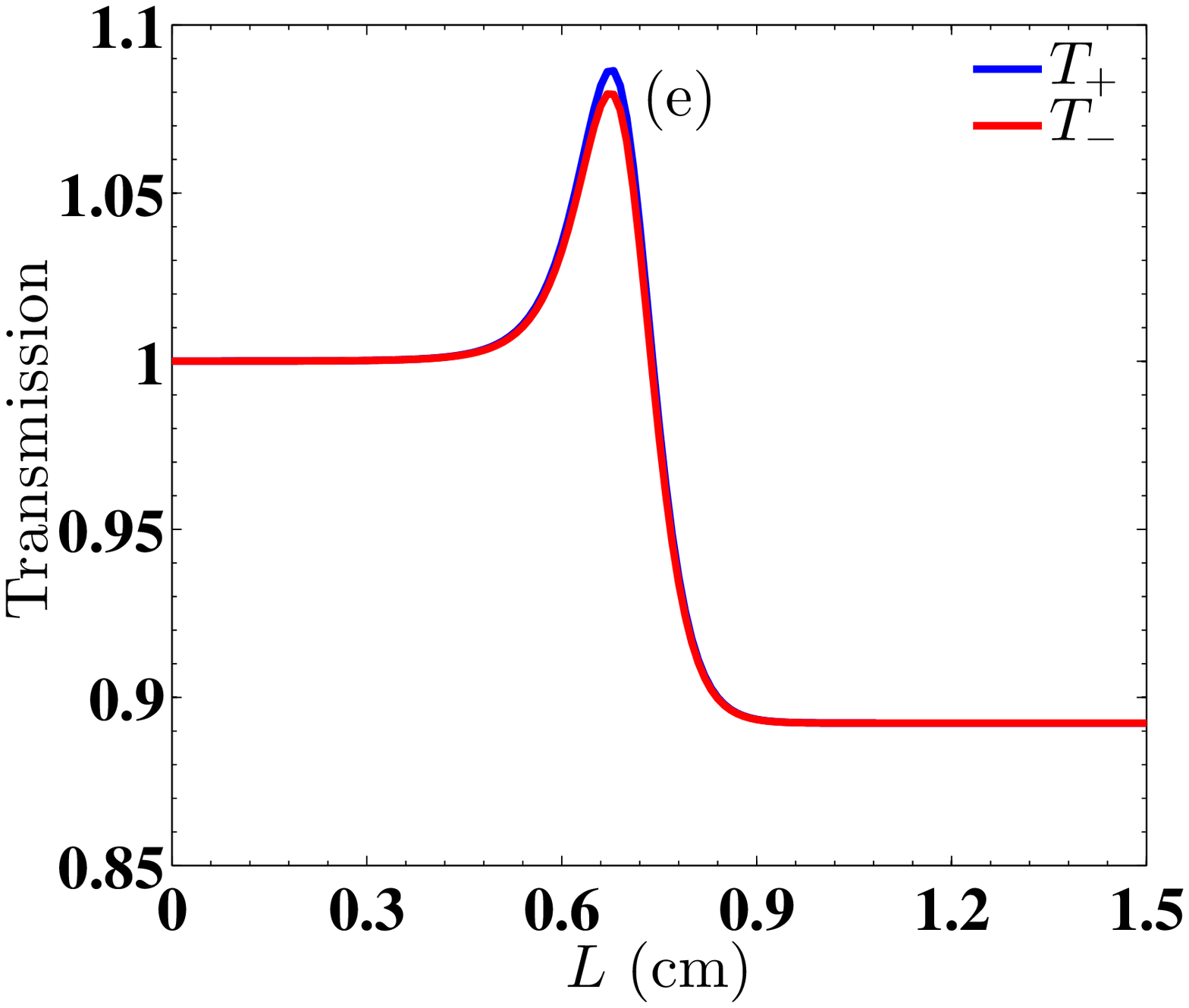}\\
\includegraphics[scale=0.23]{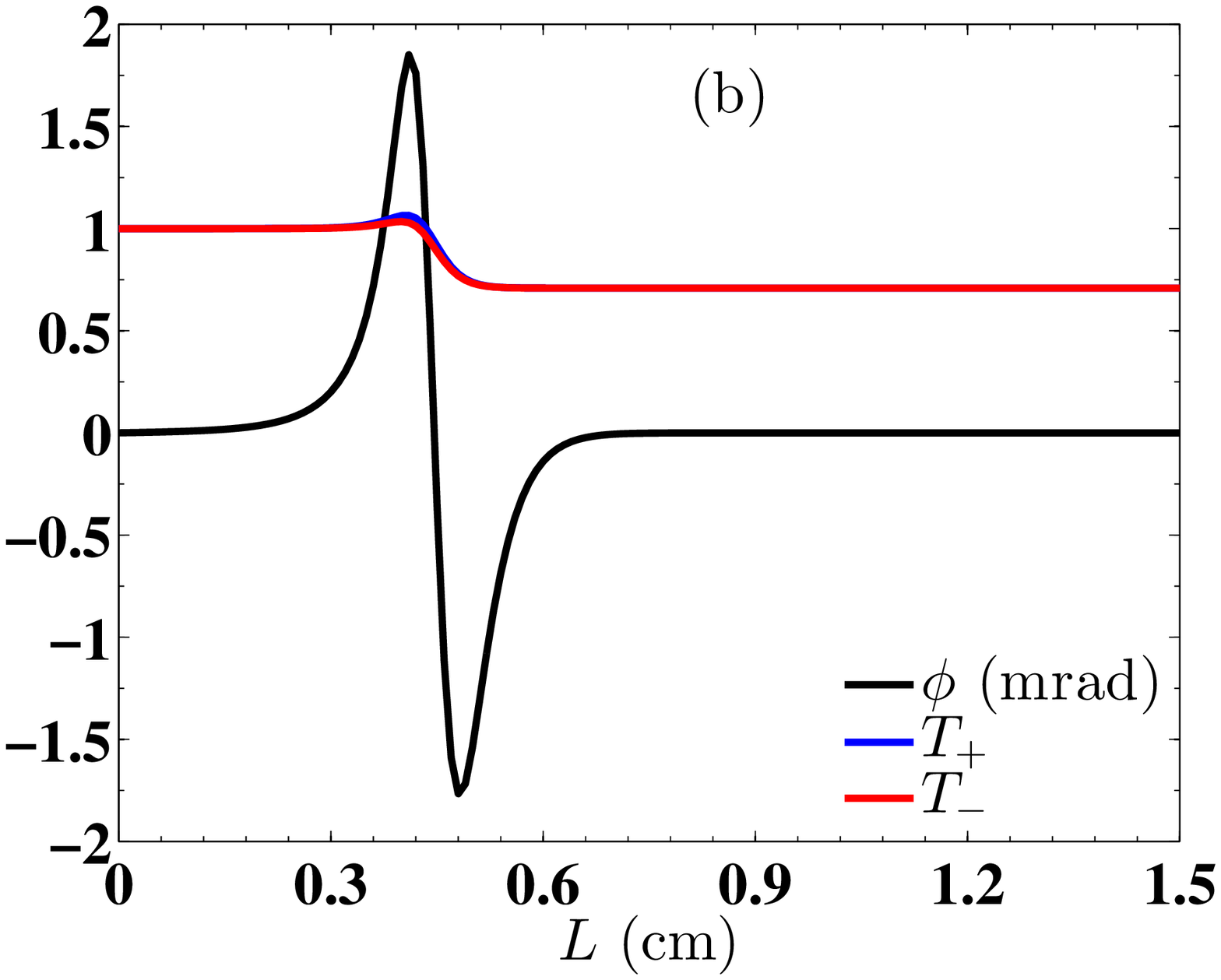}& \includegraphics[scale=0.23]{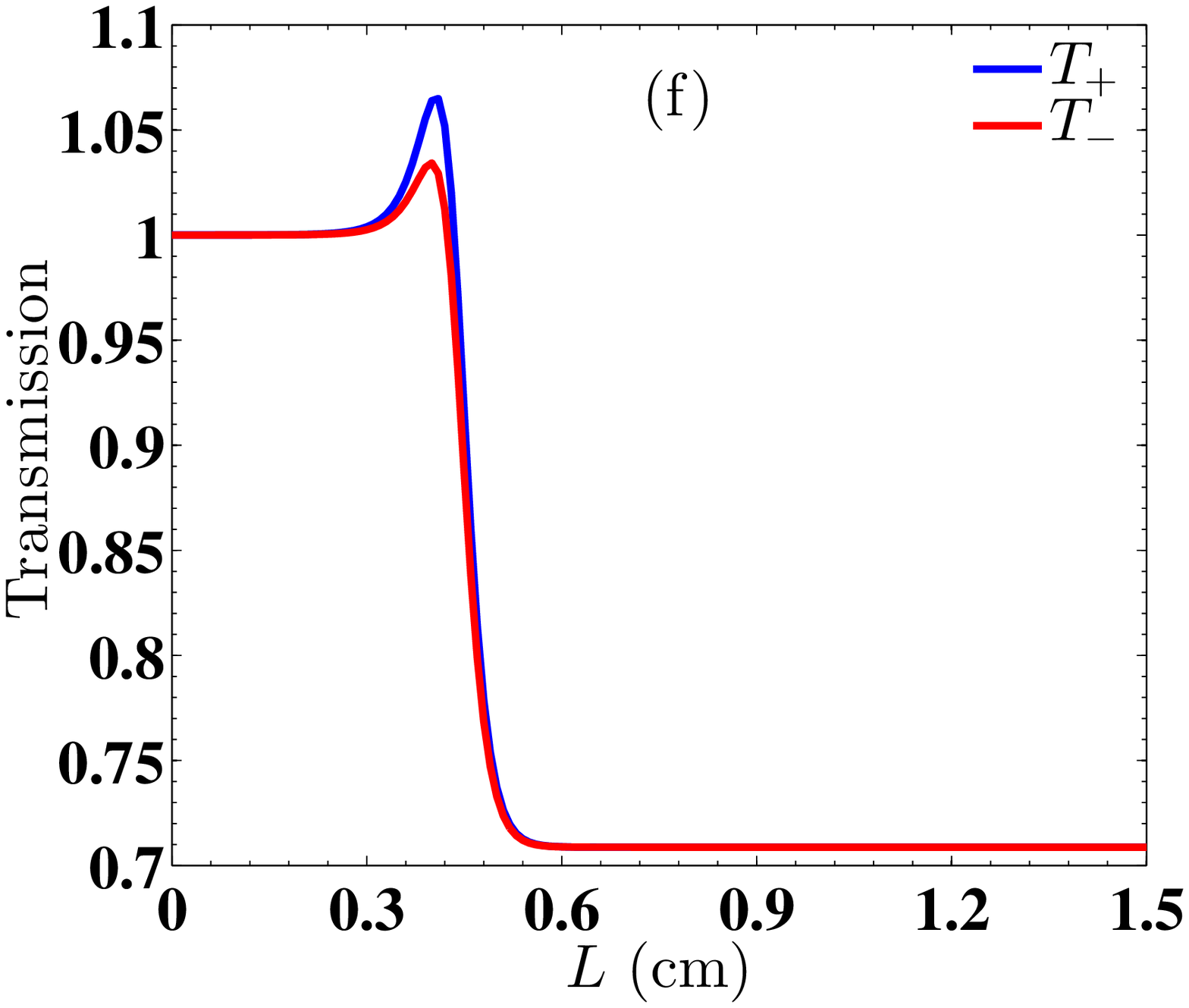}\\
\includegraphics[scale=0.23]{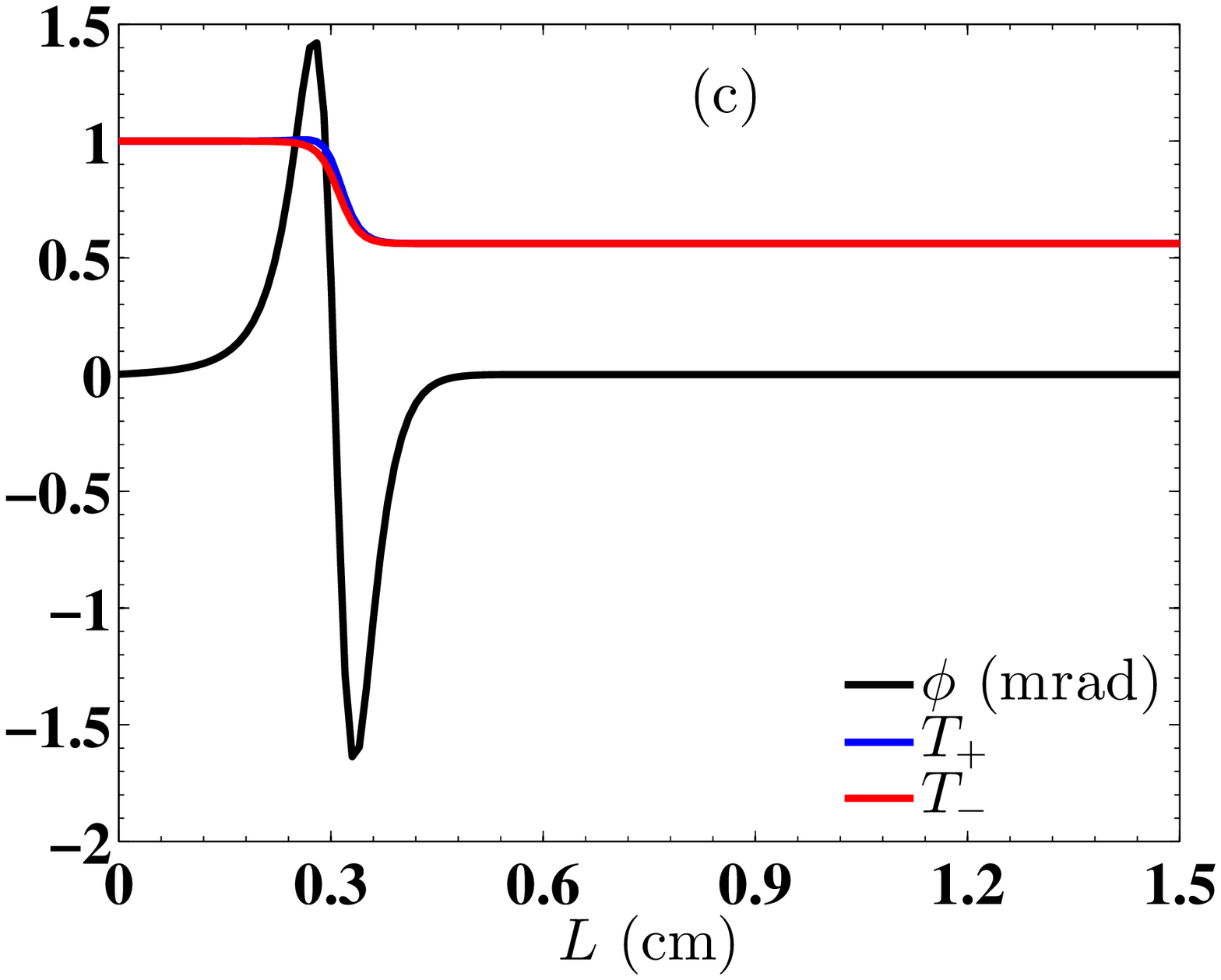}& \includegraphics[scale=0.23]{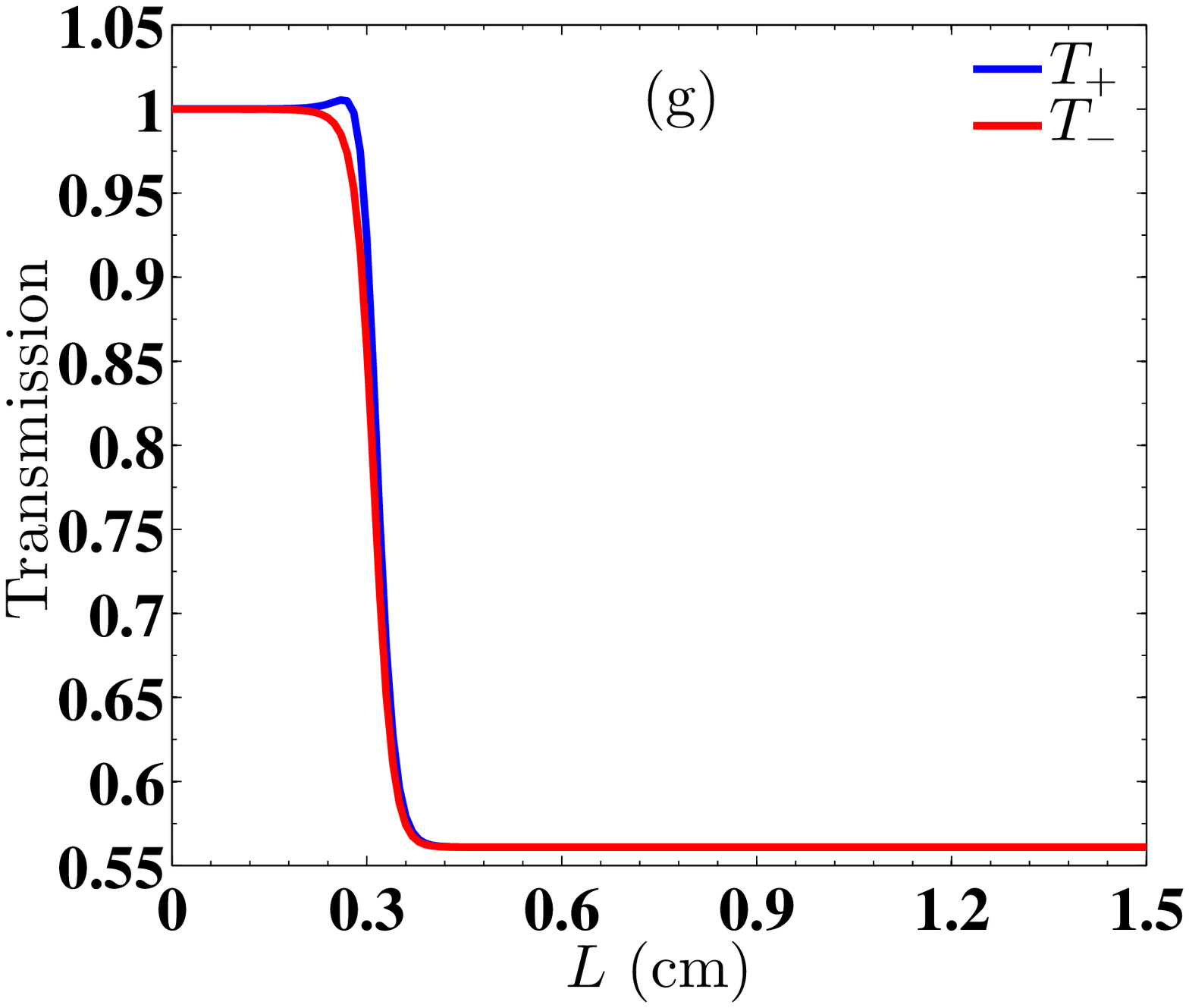}\\
\includegraphics[scale=0.23]{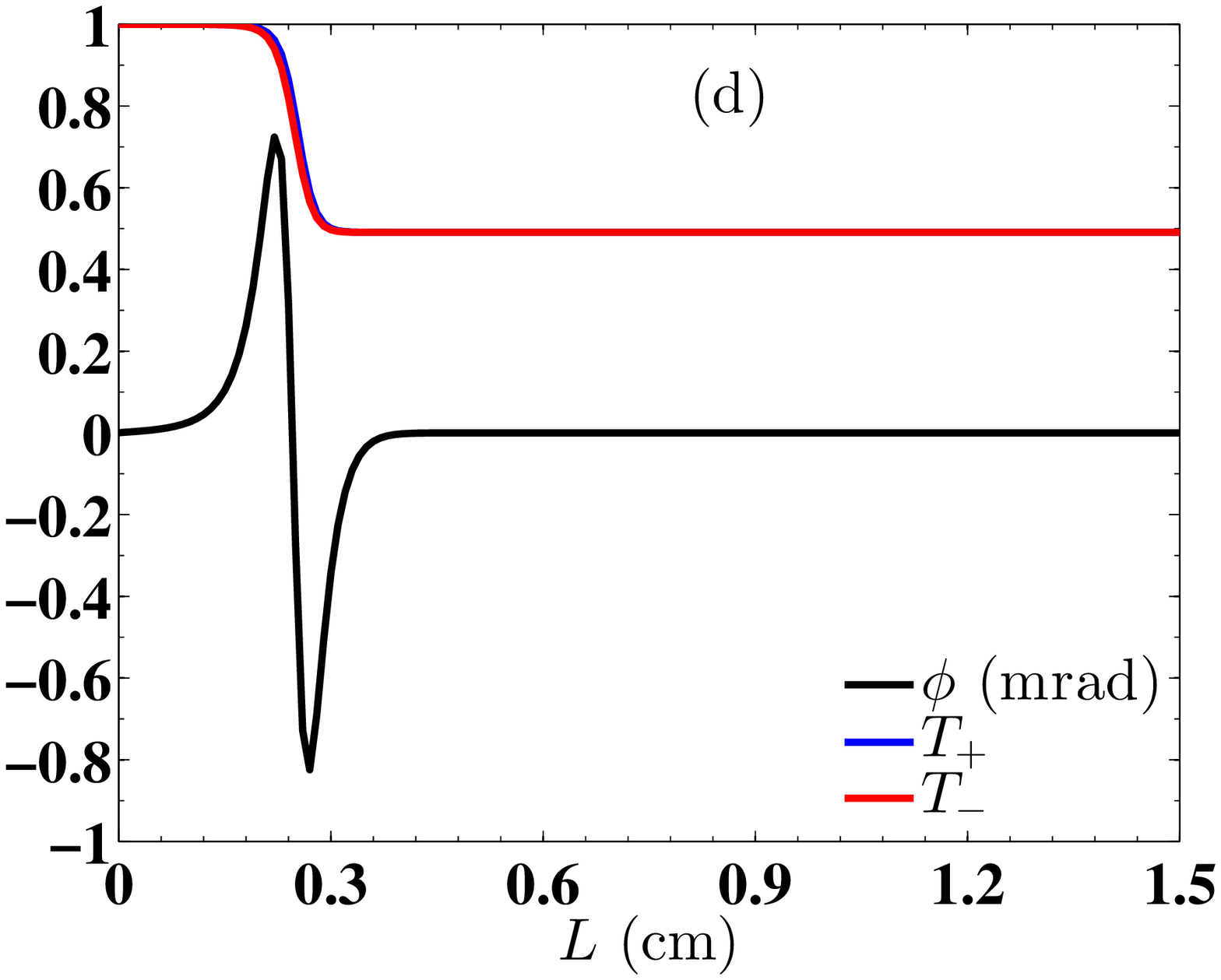}& \includegraphics[scale=0.23]{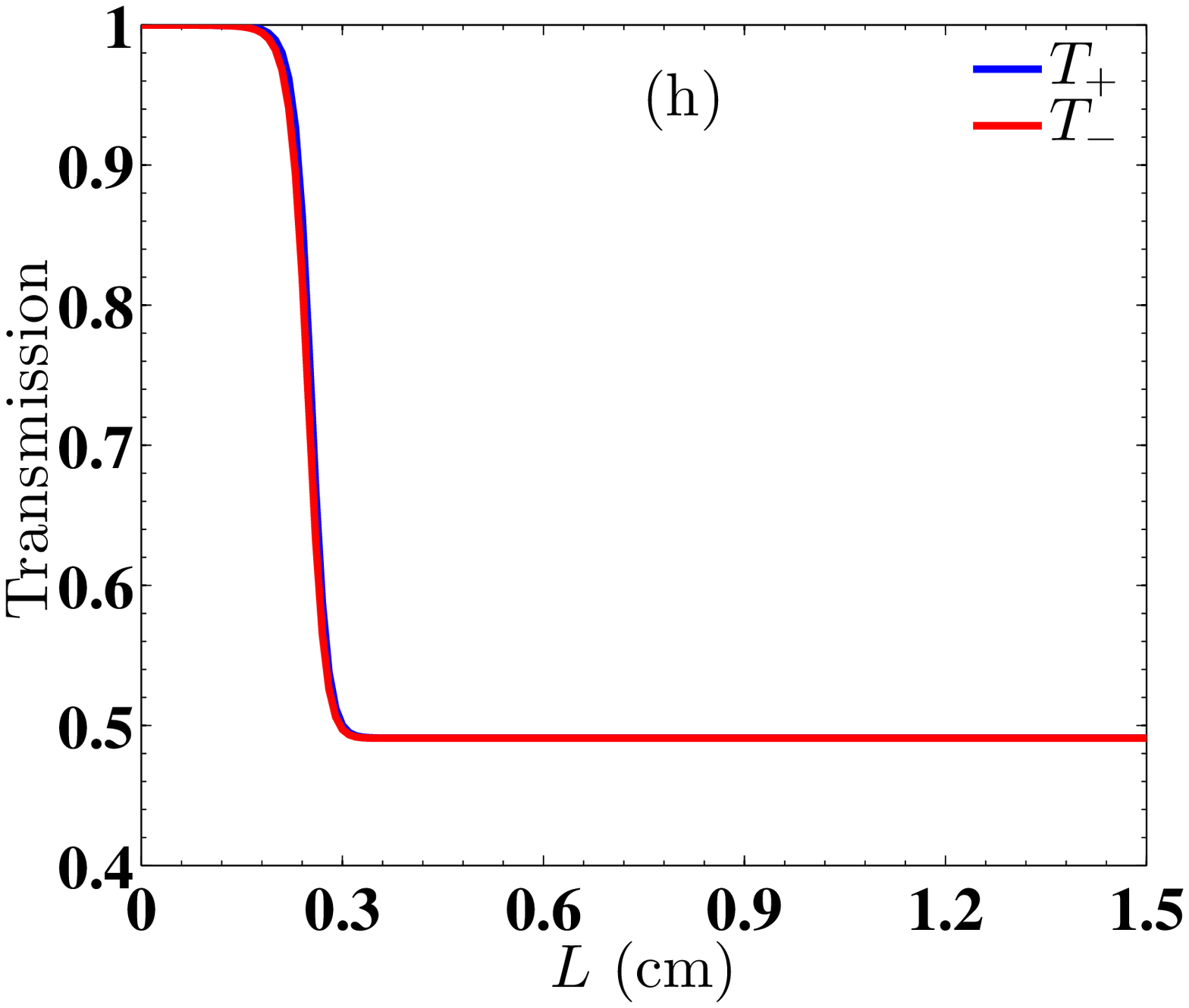}
\end{tabular}
\caption{Variation of angular divergence $\phi$ between the circular components and the respective transmission with the propagation distance $L$ (cm) for  a Gaussian profile of the control field with (a) $\theta_{c}=\pi/10$, (b) $\theta_{c}=\pi/6$, (c) $\theta_{c}=\pi/4$ and (d) $\theta_{c}=\pi/3$. The corresponding transmission is also shown in the inflated graphs (e)-(h), for the respective $\theta_c$'s. We have used the parameters for $^{39}$K vapor with $A=2\pi\times6.079$ MHz, $\gamma=\frac{A}{12}$, $\lambda=769.9$ nm, $N=5\times10^{12}$ cm$^{-3}$. Other parameters are $\Delta_{z}$=0.01$\gamma$ (32 kHz), $\delta=0$, $\Delta=0$, $\gamma_{13}=\gamma_{23}=\gamma_{03}=\gamma$, $\gamma_{coll}=0$, and the transverse width $\left(\sigma\right)$ of Gaussian profile is taken to be $\sqrt{2}$ mm.}
\label{fig2}
\end{center}
\end{figure}
To delineate the effect of angle of incidence on the angular divergence between $\sigma_{\pm}$ components of the probe field, we first consider  a Gaussian profile (LG$_{0}$) of the control field. At two-photon resonance, this  profile creates transparency for the probe field at $x$=0 whereas the position-dependent refractive index remain zero.  The circular components of the probe field experience deflection if the probe is off-centered with respect to the control field.  We demonstrate in Fig. \ref{fig2} the variation of the angular divergence between the orthogonal polarized components of the probe field along with the respective transparencies with the longitudinal distance $L$ for different values of the angle of incidence at $y=$0 plane.  The angular divergence has a dispersion like profile with maxima and minima at different axial positions, as shown in Fig. \ref{fig2}.  Note that the deflection of $\sigma_{\pm}$ components occurs within a short distance from the interface.  The maxima and minima of the angular divergence occur at a larger distance for smaller angle of incidence.  For smaller angle of incidence, the inhomogeneous control field moves close to the axis of the probe field in the vapor cell and a larger overlap area of these fields causes the angular divergence to happen at longer propagation distance [see Fig. \ref{fig2}(a)]. It is to be emphasized that during the deflection the circular components of the probe field remain transparent for the positive deflection, thanks to EIT. But, for negative deflection these components suffer absorption for the Gauusian profile of the control field as shown in Fig. \ref{fig2}(e-h).  Moreover, the absorption dominates in the system for the negative deflection as the angle of incidence of the control field increases [see Fig. \ref{fig2}(h)]. Furthermore, for larger angle of incidence [say $\theta_{c}=\pi/4$, Fig. \ref{fig2}(c)], the maximum deflection happens to be at distance  $L \approx$ FWHM of the control field.  Thus, the angle of incidence of the control field provides a flexibility to the control of the angular divergence of the polarized components of the probe field.

It is to be noted that a reasonable explanation for such a light deflection can be given in terms of the spatial dependent potential induced by the coupling between the atoms and the light \cite{zhou2008}. The transverse profile of the control field decides the shape of such a potential. Thus, a probe field of width smaller than the width of the control field gets deflected if it is adjusted to the left or to the right side of control field. Further, the deflection of the light in an EIT medium can also be described quantum mechanically with dark state \textit{polariton} possessing an effective magnetic moment \cite{karpa2006,zhou2008,zhang2009}. 
\begin{figure}[ht!]
\begin{center}
\begin{tabular}{cc}
\includegraphics[scale=0.23]{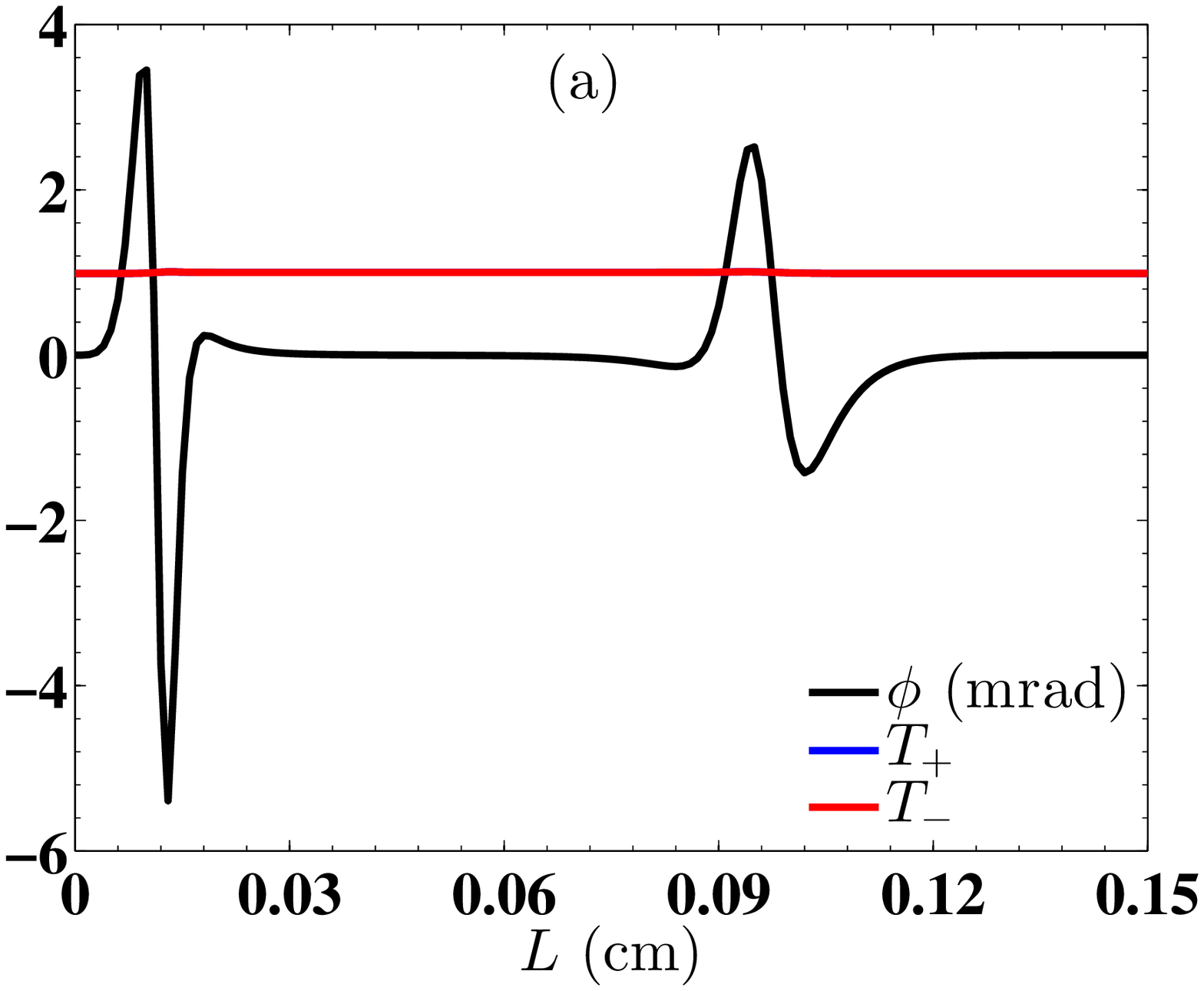}& \includegraphics[scale=0.23]{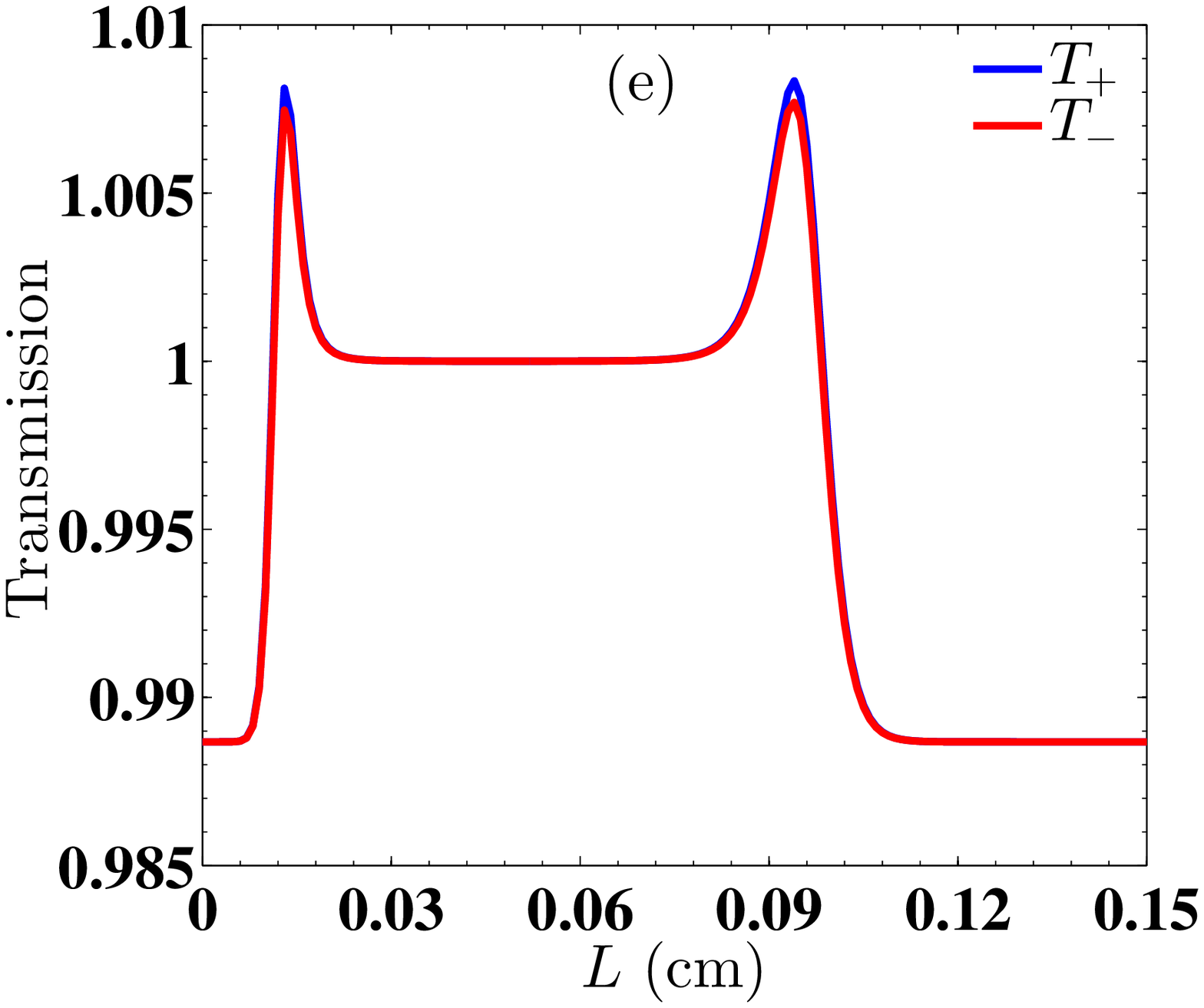}\\
\includegraphics[scale=0.23]{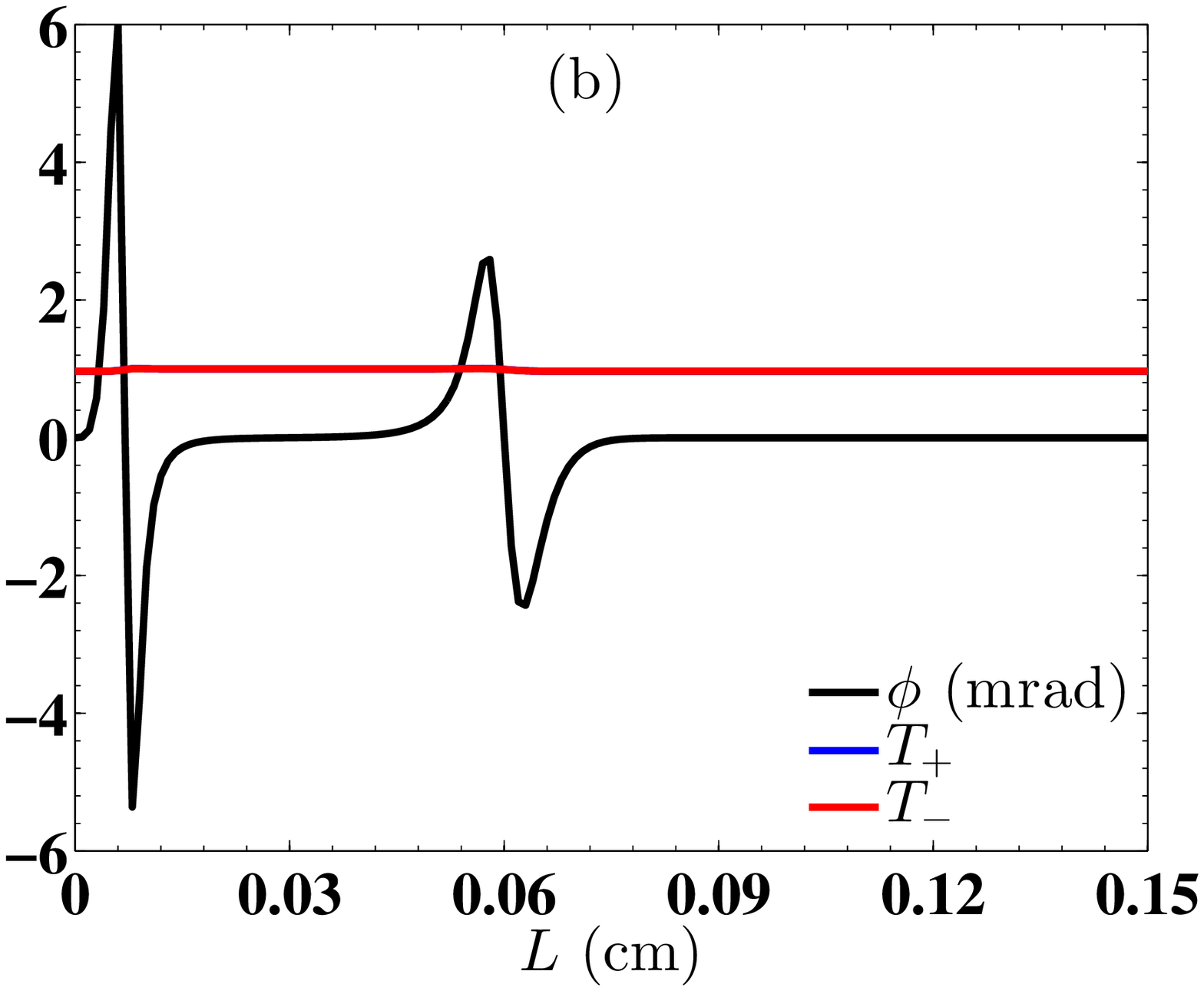}& \includegraphics[scale=0.23]{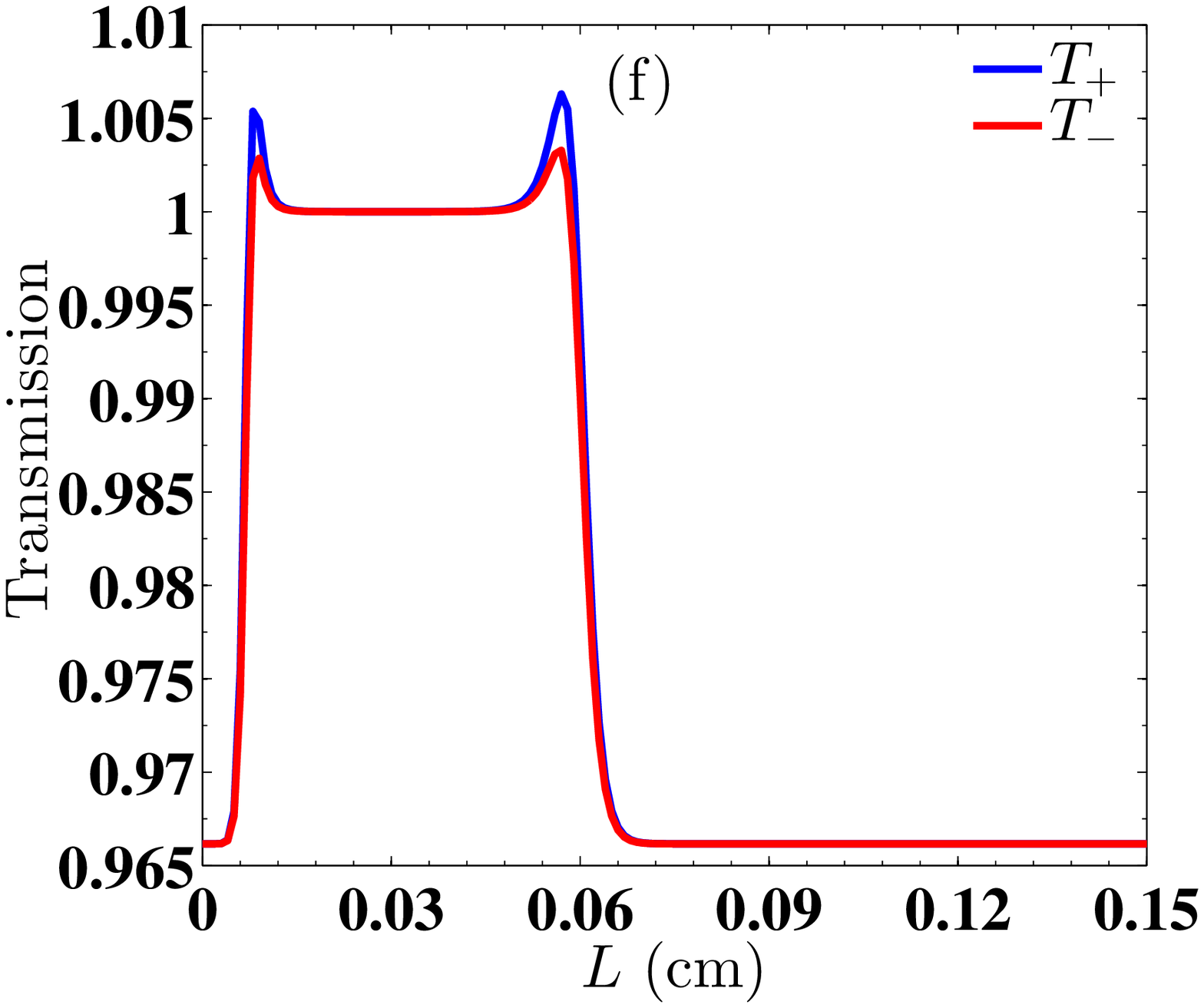}\\
\includegraphics[scale=0.23]{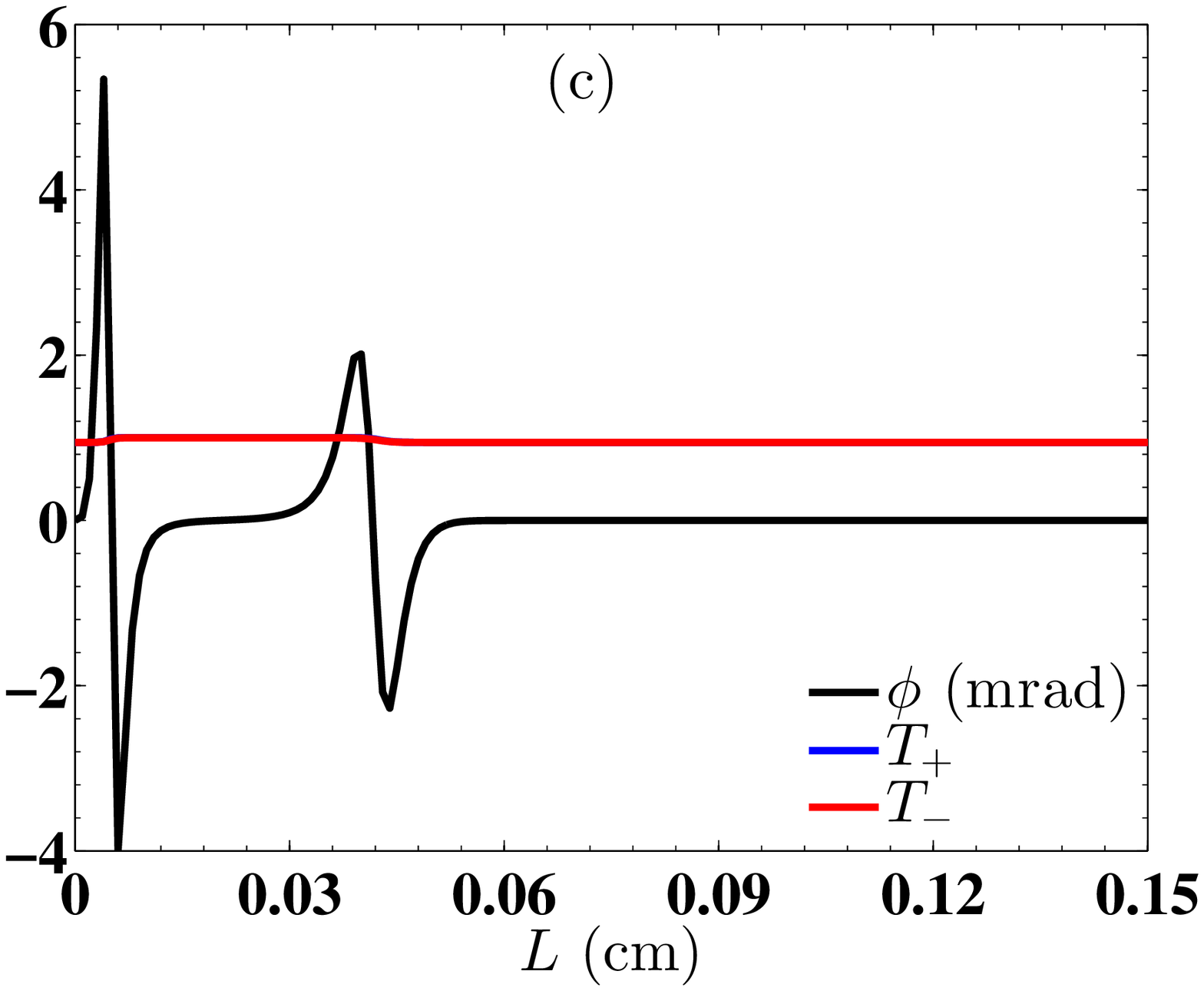}& \includegraphics[scale=0.23]{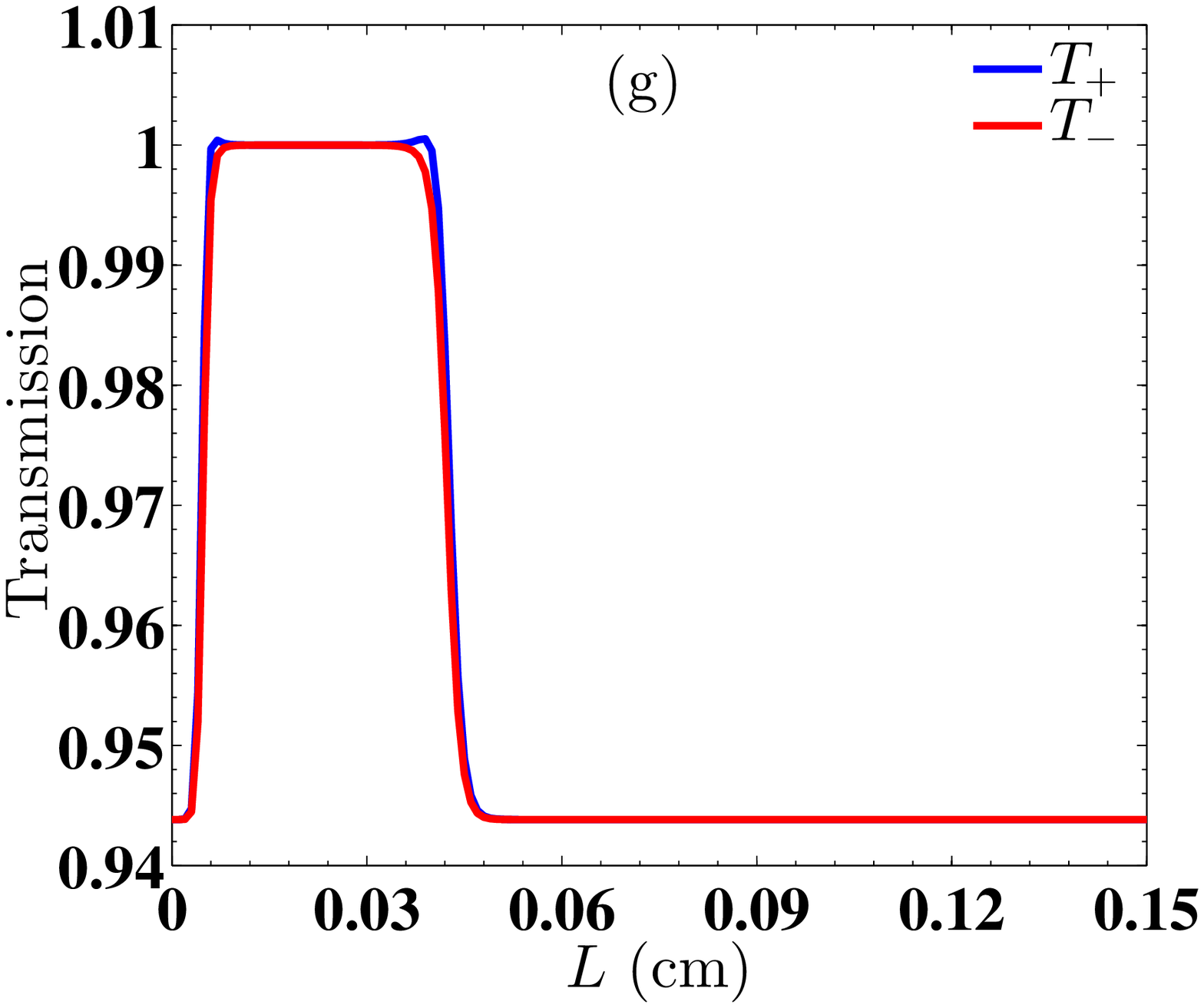}\\
\includegraphics[scale=0.23]{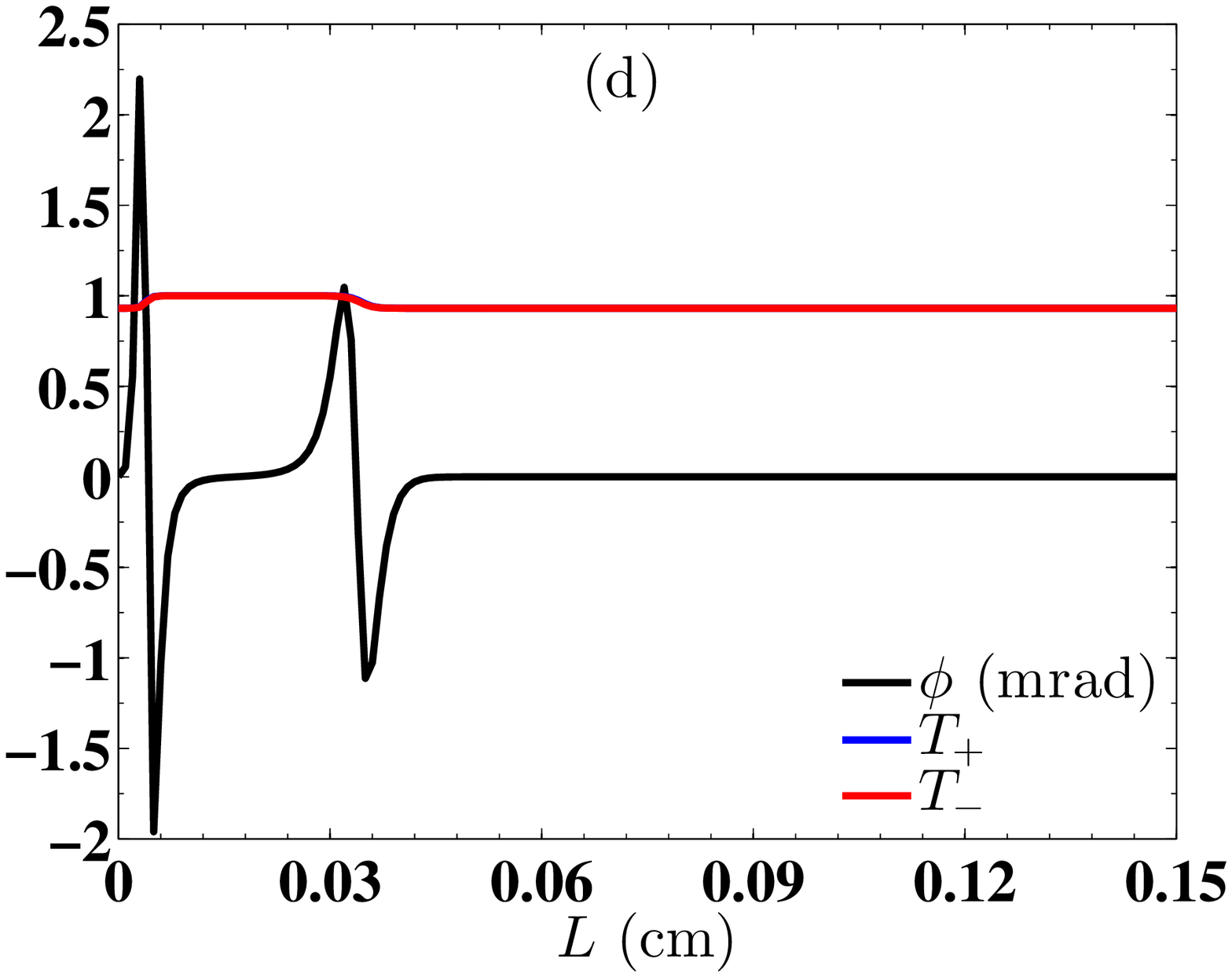}& \includegraphics[scale=0.23]{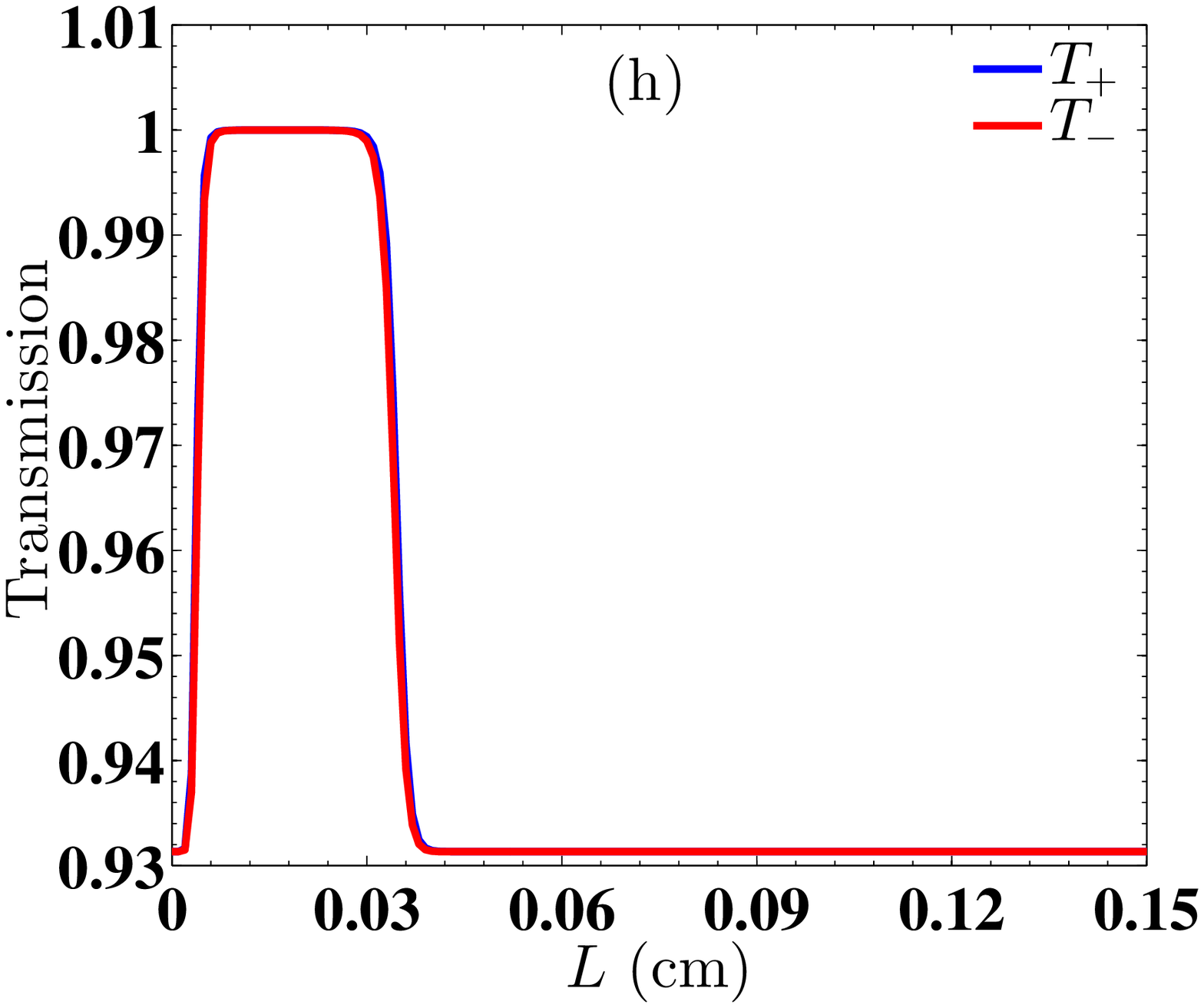}
\end{tabular}
\caption{Variation of angular divergence ($\phi$) between the circular components of the probe field and the respective transmission with the propagation distance $L$ for  a Laguerre-Gaussian profile (with $m=3$) of the control field with (a) $\theta_{c}=\pi/10$, (b) $\theta_{c}=\pi/6$, (c) $\theta_{c}=\pi/4$ and (d) $\theta_{c}=\pi/3$. The corresponding transmission is also shown in the inflated graphs (e)-(h), for the respective $\theta_c$'s. The parameters used are beam waist ($w_{0}$)=120 $\mu$m, Rayleigh length ($z_{R}$) =5.7 cm and the other parameters are the same as in Fig. \ref{fig2}.}
\label{fig3}
\end{center}
\end{figure}

\subsection{Effect of the profile of the control field}
In the preceding analysis, we have discussed how the Gaussian profile of the control field causes the deflection of the circularly polarized components of the probe field. Next, we discuss the effect of the profile of the control beam on the angular divergence between $\sigma_{\pm}$ components together with the substantial improvement of the transmission of the probe field. For this purpose, we consider a doughnut-shaped Laguerre-Gaussian mode (LG$_{3}$) for the control beam. In Fig. \ref{fig3}, we exhibit the dependence of the angular divergence and the transmission of circular components on the propagation distance for different values of the incident angles of the control field. Clearly, a comparison of Figs. \ref{fig2} and \ref{fig3} reveals that the doughnut-shaped LG$_{3}$ mode is better than the Gaussian mode to produce larger angular divergence. Also, for LG$_{3}$ transverse profile of the control field $\sigma_{\pm}$ components of the probe field remain nearly transparent throughout their propagation, as depicted in Figs. \ref{fig3}(e-h).  More interestingly, for the Laguerre-Gaussian mode, the large deflection occurs without much absorption within a very short distance from the entrance face of the vapor cell. This feature mimics the refraction of a linearly polarized light from the interface, as in the case of a medium with chiral molecules \cite{ghosh2006}. For a medium having natural anisotropy due to the presence of chiral molecules, a linearly polarized incident light splits into circularly polarized components just at the interface. Here, we have shown such an effect in simple atomic system.
Furthermore, the location of maximum angular divergence of the orthogonal components close to the entry face of the atomic vapor cell can be modified by changing the beam waist. Increase in the beam waist causes the deflection to happen at longer distance. 
\subsection{Lens effects of a coherently prepared atomic medium}
\begin{figure}[ht!]
\begin{center}
\begin{tabular}{c}
\includegraphics[scale=0.35]{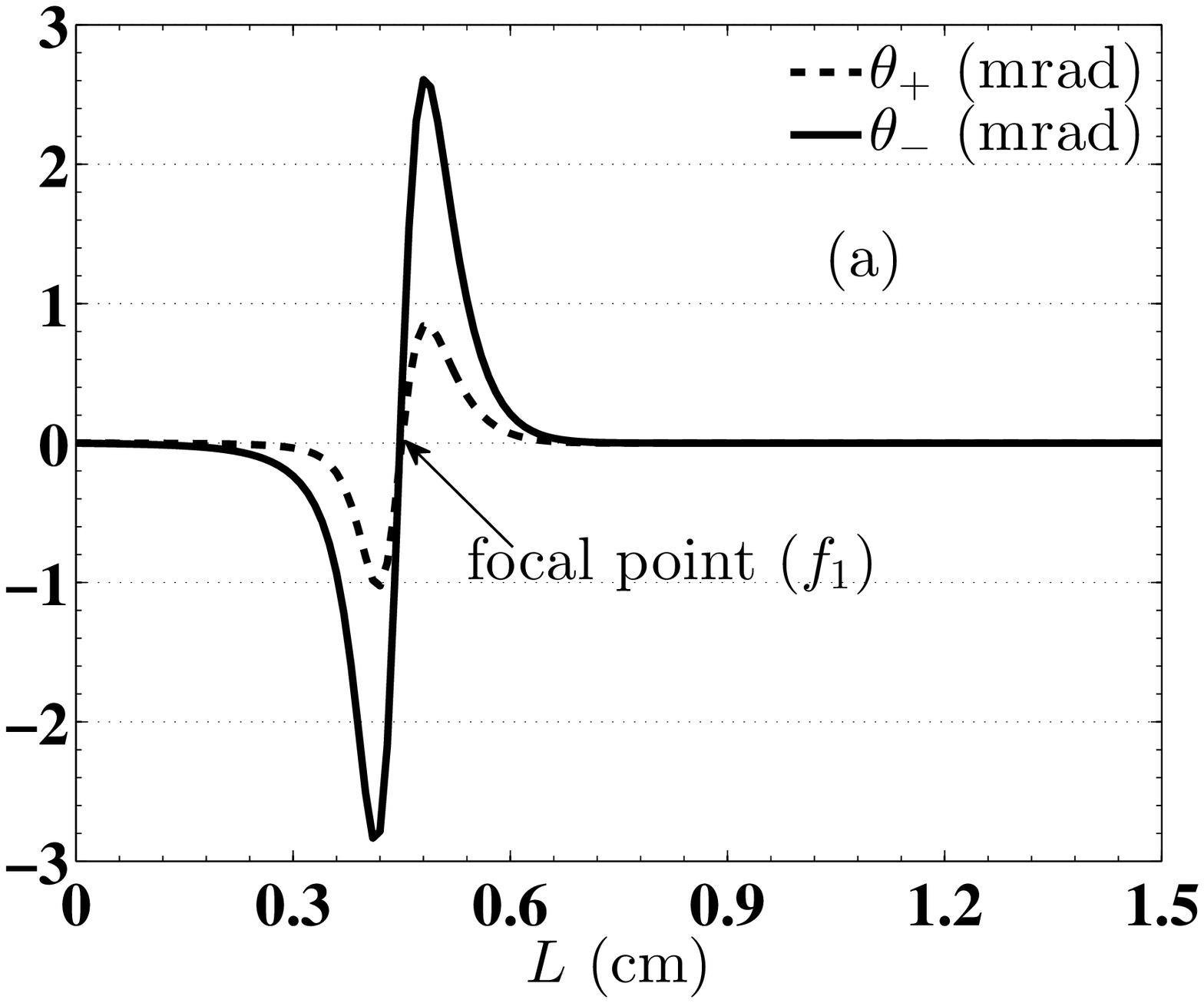}\\ \includegraphics[scale=0.35]{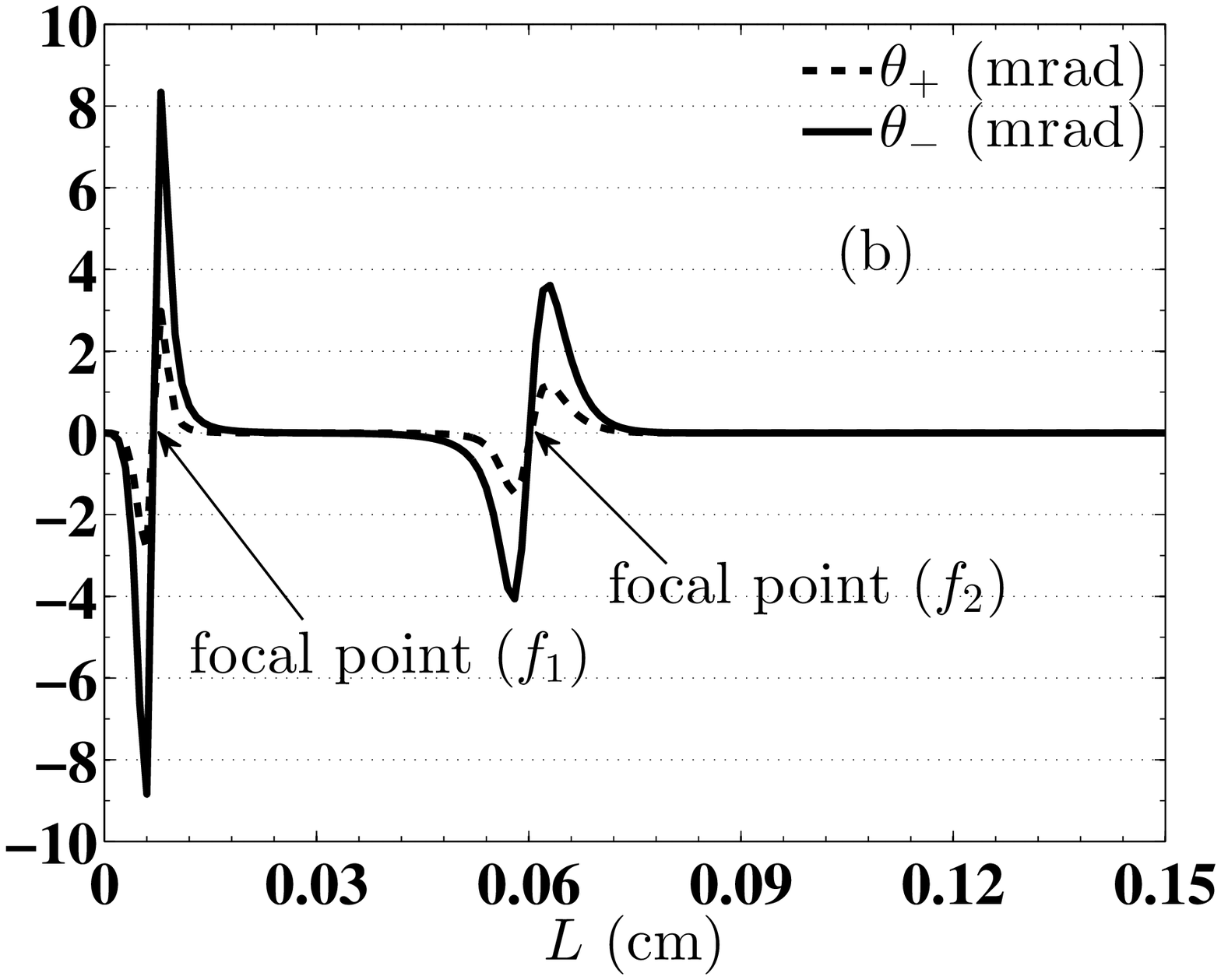}
\end{tabular}
\caption{The deflection of $\sigma_{\pm}$ components of the probe field with the propagation length $L$ for (a) the Gaussian (b) LG$_{3}$ profile of the control field. We have chosen $\theta_{c}=\pi/6$ and the rest of the parameters are the same as in Fig. \ref{fig2}.}
\label{fig4}
\end{center}
\end{figure}
The above analysis implies the deflection of the orthogonally polarized components of the probe field through a coherently prepared medium. Now, we demonstrate that such a medium behaves as a lens with a varying focus which can be controlled optically. Such a lens causes the focusing and refocusing of $\sigma_{\pm}$ components. This is demonstrated in Fig. \ref{fig4}. For the Gaussian profile of the control beam [Fig. \ref{fig4}(a)], the two orthogonal circular components converge at a point $f_{1}$ which refers to a focal point.  On the other hand, for LG$_{3}$ transverse profile of the control field [Fig. \ref{fig4} (b)], $\sigma_{\pm}$ components refocus at $f_{2}$ after focusing at $f_{1}$, thereby producing two focal points. Thus, a coherent medium behaves as a converging lens, the number of focal points of which can be controlled optically. Further, the location of the focal points can be varied by changing the angle of incidence of the control field. Thus, the profile of the control beam and its angle of incidence act as a knob to control the focusing and refocusing of the probe field while passing through an optically controlled atomic medium.
\section{Conclusions}
In conclusion, we have explored the enhancement of the angular divergence between the orthogonally polarized components of the linearly polarized probe field by using angle of incidence and the profile of the control beam as a knob. Such an effect can be enhanced by using higher order Laguerre-Gaussian modes in the profile of the control field. For LG$_{3}$ mode, the maximum deflection happens close to the entry face of the vapor cell. Moreover, a large angular divergence close to entry face is accompanied by negligible absorption due to EIT.  We further show how the coherent atomic medium can act as a lens with multiple foci, depending upon the profile and the angle of incidence of the control beam.
\appendix*
\section{Relevant Density matrix equations}
To describe the dynamics of the system, we use Markovian master equation under rotating wave approximation by including the natural decay terms  and following density matrix equations are obtained:
\begin{eqnarray}
\begin{array}{lll}
\dot{\tilde{\rho}}_{11} &=&\gamma_{13}\tilde{\rho}_{33}-i\left(g_{1}e^{-i\omega_{pc}t}\tilde{\rho}_{13}-c.c.\right)-i\left(G_{1}\tilde{\rho}_{13}-c.c.\right)\;,\\
\dot{\tilde{\rho}}_{22} &=&\gamma_{23}\tilde{\rho}_{33}-i\left(g_{2}e^{-i\omega_{pc}t}\tilde{\rho}_{23}-c.c.\right)-i\left(G_{2}\tilde{\rho}_{23}-c.c.\right)\;,\\
\dot{\tilde{\rho}}_{00} &=&\gamma_{03}\tilde{\rho}_{33}-i\left(G_{0}\tilde{\rho}_{03}-c.c.\right)\;,\\

\dot{\tilde{\rho}}_{31}&=&\left[i\left(\Delta+\Delta_{z}\right)-\Gamma_{31}\right]\tilde{\rho}_{31}+iG_{2}\tilde{\rho}_{21}+iG_{0}\tilde{\rho}_{01}\\
&&+ig_{2}e^{-i\omega_{pc}t}\tilde{\rho}_{21}+i\left[g_{1}e^{-i\omega_{pc}t}+G_{1}\right]\left(\tilde{\rho}_{11}-\tilde{\rho}_{33}\right)\;,\\

\dot{\tilde{\rho}}_{32} &=&\left[i\left(\Delta-\Delta_{z}\right)-\Gamma_{32}\right]\tilde{\rho}_{32}+ig_{1}e^{-i\omega_{pc}t}\tilde{\rho}_{12}+iG_{0}\tilde{\rho}_{02}\\
&&+iG_{1}\tilde{\rho}_{12}+i\left[g_{2}e^{-i\omega_{pc}t}+G_{2}\right]\left(\tilde{\rho}_{22}-\tilde{\rho}_{33}\right)\;,\\

\dot{\tilde{\rho}}_{30} &=&\left[i\Delta-\Gamma_{30}\right]\tilde{\rho}_{30}+ie^{-i\omega_{pc}t}\left(g_{1}\tilde{\rho}_{10}+g_{2}\tilde{\rho}_{20}\right)\\
&&+i\left(G_{1}\tilde{\rho}_{10}+G_{2}\tilde{\rho}_{20}\right)+iG_{0}\left(\tilde{\rho}_{00}-\tilde{\rho}_{33}\right)\;,\\

\dot{\tilde{\rho}}_{01} &=&\left[i\Delta_{z}-\Gamma_{01}\right]\tilde{\rho}_{01}+iG^{\ast}_{0}\tilde{\rho}_{31}-i\left[g_{1}e^{-i\omega_{pc}t}+G^{\ast}_{1}\right]\tilde{\rho}_{03}\;,\\

\dot{\tilde{\rho}}_{20} &=&-\left[i\Delta_{z}+\Gamma_{20}\right]\tilde{\rho}_{20}-iG^{\ast}_{0}\tilde{\rho}_{23}+i\left[g^{\ast}_{2}e^{i\omega_{pc}t}+G^{\ast}_{2}\right]\tilde{\rho}_{30}\;,\\

\dot{\tilde{\rho}}_{21} &=&-\left[2i\Delta_{z}+\Gamma_{21}\right]\tilde{\rho}_{21}-i\left[g_{1}e^{-i\omega_{pc}t}+G^{\ast}_{1}\right]\tilde{\rho}_{23}\\
&&+i\left[g^{\ast}_{2}e^{i\omega_{pc}t}+G^{\ast}_{2}\right]\tilde{\rho}_{31}\;.\\
\end{array}
\label{eqa1}
\end{eqnarray}

The above density matrix equations are subjected to the conditions $\sum\limits_{i}\tilde{\rho}_{ii}=1$ and $\tilde{\rho}_{ij}=\tilde{\rho}_{ji}^{*}$. Here , $\Delta=\omega_{c}-\omega_{30}~\left(\delta=\omega_{p}-\omega_{30}\right)$ is the detuning of the control field (probe field) from the transition $|0\rangle\leftrightarrow|3\rangle$ transition and $\omega_{pc}=\delta-\Delta$ is the frequency difference between probe and control field. Here, $\gamma_{ij}$ is the spontaneous emission rate from the level $|j\rangle$ to $|i\rangle$, $\Gamma_{ij}=\frac{1}{2}\sum\limits_{k}\left(\gamma_{ki}+\gamma_{kj}\right)+\gamma_{coll}$ is the dephasing rate of the coherence between the levels $|j\rangle$ and $|i\rangle$, and $\gamma_{coll}$ is the collisional decay rate. The transformations for the density matrix elements are as follows: $\rho_{3j}=\tilde{\rho}_{3j}e^{-i\omega_{c}t}~\left(j=0,1,2\right)$ and the rest of the elements remain the same.
In the weak probe field limit, the density matrix elements can be expanded to first order in $g$'s in terms of the harmonics $\omega_{pc}$ as
\begin{align}\label{eqa2}
\tilde{\rho}_{\alpha\beta}&=\tilde{\rho}^{(0)}_{\alpha\beta}+g_{1}e^{-i\omega_{pc}t}\tilde{\rho}^{\prime (+1)}_{\alpha\beta}+g^{\ast}_{1}e^{i\omega_{pc}t}\tilde{\rho}^{\prime \prime(+1)}_{\alpha\beta}\nonumber\\
&+g_{2}e^{-i\omega_{pc}t}\tilde{\rho}^{\prime (-1)}_{\alpha\beta}+g^{\ast}_{2}e^{i\omega_{pc}t}\tilde{\rho}^{\prime \prime(-1)}_{\alpha\beta}\;.
\end{align}
where $\tilde{\rho}_{\alpha\beta}^{(0)}$ represents the zeorth order solution in the absence of the probe field and $\tilde{\rho}_{\alpha\beta}^{(n)}$ describes the $n$th-order solution. By substituting Eq. (\ref{eqa2}) in Eq. (\ref{eqa1}) and equating the like terms, we obtain the following algebraic equations for the zeroth and first order coherences:
\begin{equation}
A_{k}X_{k}=B_{k}~~~~~~~\left(k=0,\pm\right)\;,
\label{eqa3}
\end{equation}
The explicit form of the various terms in Eq. (\ref{eqa3}) can be written as follows:
\begin{widetext}
\begin{align}\label{eqa4}
A_{0} &=\left[\begin{array}{ccccccccccccccc}
-\gamma_{13} & -\gamma_{13} & -\gamma_{13} & \Theta_{1c}^{\ast} & \Theta_{1c} & 0 & 0 & 0 & 0 & 0 & 0 & 0 & 0 & 0 & 0\\
-\gamma_{23} & -\gamma_{23} & -\gamma_{23} & 0 & 0 & \Theta_{2c}^{\ast} & \Theta_{2c} & 0 & 0 & 0 & 0 & 0 & 0 & 0 & 0\\
-\gamma_{03} & -\gamma_{03} & -\gamma_{03} & 0 & 0 & 0 & 0 & \Phi_{0c}^{\ast} & \Phi_{0c} & 0 & 0 & 0 & 0 & 0 & 0 \\
-2\Theta_{1c} & -\Theta_{1c} & -\Theta_{1c} & p_{31} & 0 & 0 & 0 & 0 & 0 & -\Phi_{0c} & 0 & 0 & 0 & -\Theta_{2c} & 0\\
-2\Theta_{1c}^{\ast} & -\Theta_{1c}^{\ast} & -\Theta_{1c}^{\ast} & 0 & p_{31}^{\ast} & 0 & 0 & 0 & 0 & 0 & -\Phi_{0c}^{\ast} & 0 & 0 & 0 & -\Theta_{2c}^{\ast}\\
-\Theta_{2c} & -2\Theta_{2c} & -\Theta_{2c} & 0 & 0 & p_{32} & 0 & 0 & 0 & 0 & 0 & 0 & -\Phi_{0c} & 0 & -\Theta_{1c} \\
-\Theta_{2c}^{\ast} & -2\Theta_{2c}^{\ast} & -\Theta_{2c}^{\ast} & 0 & 0 & 0 & p_{32}^{\ast} & 0 & 0 & 0 & 0 & -\Phi_{0c}^{\ast} & 0 & -\Theta_{1c}^{\ast} & 0 \\
-\Phi_{0c} & -\Phi_{0c} & -2\Phi_{0c} & 0 & 0 & 0 & 0 & p_{30} & 0 & 0 & -\Theta_{1c} & -\Theta_{2c} & 0 & 0 & 0\\
-\Phi_{0c}^{\ast} & -\Phi_{0c}^{\ast} & -2\Phi_{0c}^{\ast} & 0 & 0 & 0 & 0 & 0 & p_{30}^{\ast} & -\Theta_{1c}^{\ast} & 0 & 0 & -\Theta_{2c}^{\ast} & 0 & 0 \\
0 & 0 & 0 & \Phi_{0c}^{\ast} & 0 & 0 & 0 & 0 & \Theta_{1c} & p_{01} & 0 & 0 & 0 & 0 & 0\\
0 & 0 & 0 & 0 & \Phi_{0c} & 0 & 0 & \Theta_{1c}^{\ast} & 0 & 0 & p_{01}^{\ast} & 0 & 0 & 0 & 0\\
0 & 0 & 0 & 0 & 0 & 0 & \Phi_{0c} & \Theta_{2c}^{\ast} & 0 & 0 & 0 & p_{20} & 0 & 0 & 0\\
0 & 0 & 0 & 0 & 0 & \Phi_{0c}^{\ast} & 0 & 0 & \Theta_{2c} & 0 & 0 & 0 & p_{20}^{\ast} & 0 & 0\\
0 & 0 & 0 & \Theta_{2c}^{\ast} & 0 & 0 & \Theta_{1c} & 0 & 0 & 0 & 0 & 0 & 0 & p_{21} & 0\\
0 & 0 & 0 & 0 & \Theta_{2c} & \Theta_{1c}^{\ast} & 0 & 0 & 0 & 0 & 0 & 0 & 0 & 0 & p_{21}^{\ast}
\end{array}\right]\;,\\
X_{0}&=\left[\begin{array}{lllllllllllllll}
\tilde{\rho}_{11}^{(0)} & \tilde{\rho}_{22}^{(0)} & \tilde{\rho}_{00}^{(0)} & \tilde{\rho}_{31}^{(0)} & \tilde{\rho}_{13}^{(0)} & \tilde{\rho}_{32}^{(0)} & \tilde{\rho}_{23}^{(0)} & \tilde{\rho}_{30}^{(0)} & \tilde{\rho}_{03}^{(0)} & \tilde{\rho}_{01}^{(0)} & \tilde{\rho}_{10}^{(0)} & \tilde{\rho}_{20}^{(0)} & \tilde{\rho}_{02}^{(0)} & \tilde{\rho}_{21}^{(0)} & \tilde{\rho}_{12}^{(0)}
\end{array}\right]^{T}\;,\label{eqa5}\\
B_{0}&=\left[\begin{array}{lllllllllllllll}
-\gamma_{13} & -\gamma_{23} & -\gamma_{03} & \Theta_{1c} & -\Theta_{1c}^{\ast} & \Theta_{2c} & -\Theta_{2c}^{\ast} & \Phi_{0c} & -\Phi_{0c}^{\ast} & 0 & 0 & 0 & 0 & 0 & 0
\end{array}\right]^{T}\;.\label{eqa6}
\end{align}
where $p_{31}=i\left(\Delta+\Delta_{z}\right)-\Gamma_{31}$,  $p_{32}=i\left(\Delta-\Delta_{z}\right)-\Gamma_{32}$,  $p_{30}=i\Delta-\Gamma_{30}$, $p_{01}=i\Delta_{z}-\Gamma_{01}$,  $p_{20}=-i\Delta_{z}-\Gamma_{20}$, $p_{21}=-2i\Delta_{z}-\Gamma_{21}$,  $\Theta_{1c}=-iG_{1}$, $\Theta_{2c}=-iG_{2}$,  $\Phi_{0c}=-iG_{0}$, and $T$ denotes the the transpose of the vector.
\begin{align}\label{eqa7}
A_{\pm} &=
\left[\begin{array}{ccccccccccccccc}
p_{11}^{\prime} & -\gamma_{13} & -\gamma_{13} & \Theta_{1c}^{\ast} & \Theta_{1c} & 0 & 0 & 0 & 0 & 0 & 0 & 0 & 0 & 0 & 0\\
-\gamma_{23} & p_{22}^{\prime} & -\gamma_{23} & 0 & 0 & \Theta_{2c}^{\ast} & \Theta_{2c} & 0 & 0 & 0 & 0 & 0 & 0 & 0 & 0\\
-\gamma_{03} & -\gamma_{03} & p_{00}^{\prime} & 0 & 0 & 0 & 0 & \Phi_{0c}^{\ast} & \Phi_{0c} & 0 & 0 & 0 & 0 & 0 & 0 \\
-2\Theta_{1c} & -\Theta_{1c} & -\Theta_{1c} & p_{31}^{\prime} & 0 & 0 & 0 & 0 & 0 & -\Phi_{0c} & 0 & 0 & 0 & -\Theta_{2c} & 0\\
-2\Theta_{1c}^{\ast} & -\Theta_{1c}^{\ast} & -\Theta_{1c}^{\ast} & 0 & p_{13}^{\prime} & 0 & 0 & 0 & 0 & 0 & -\Phi_{0c}^{\ast} & 0 & 0 & 0 & -\Theta_{2c}^{\ast}\\
-\Theta_{2c} & -2\Theta_{2c} & -\Theta_{2c} & 0 & 0 & p_{32}^{\prime} & 0 & 0 & 0 & 0 & 0 & 0 & -\Phi_{0c} & 0 & -\Theta_{1c} \\
-\Theta_{2c}^{\ast} & -2\Theta_{2c}^{\ast} & -\Theta_{2c}^{\ast} & 0 & 0 & 0 & p_{23}^{\prime} & 0 & 0 & 0 & 0 & -\Phi_{0c}^{\ast} & 0 & -\Theta_{1c}^{\ast} & 0 \\
-\Phi_{0c} & -\Phi_{0c} & -2\Phi_{0c} & 0 & 0 & 0 & 0 & p_{30}^{\prime} & 0 & 0 & -\Theta_{1c} & -\Theta_{2c} & 0 & 0 & 0\\
-\Phi_{0c}^{\ast} & -\Phi_{0c}^{\ast} & -2\Phi_{0c}^{\ast} & 0 & 0 & 0 & 0 & 0 & p_{03}^{\prime} & -\Theta_{1c}^{\ast} & 0 & 0 & -\Theta_{2c}^{\ast} & 0 & 0 \\
0 & 0 & 0 & \Phi_{0c}^{\ast} & 0 & 0 & 0 & 0 & \Theta_{1c} & p_{01}^{\prime} & 0 & 0 & 0 & 0 & 0\\
0 & 0 & 0 & 0 & \Phi_{0c} & 0 & 0 & \Theta_{1c}^{\ast} & 0 & 0 & p_{10}^{\prime} & 0 & 0 & 0 & 0\\
0 & 0 & 0 & 0 & 0 & 0 & \Phi_{0c} & \Theta_{2c}^{\ast} & 0 & 0 & 0 & p_{20}^{\prime} & 0 & 0 & 0\\
0 & 0 & 0 & 0 & 0 & \Phi_{0c}^{\ast} & 0 & 0 & \Theta_{2c} & 0 & 0 & 0 & p_{02}^{\prime} & 0 & 0\\
0 & 0 & 0 & \Theta_{2c}^{\ast} & 0 & 0 & \Theta_{1c} & 0 & 0 & 0 & 0 & 0 & 0 & p_{21}^{\prime} & 0\\
0 & 0 & 0 & 0 & \Theta_{2c} & \Theta_{1c}^{\ast} & 0 & 0 & 0 & 0 & 0 & 0 & 0 & 0 & p_{12}^{\prime}
\end{array}\right]\;,\\
X_{\pm} &=\left[\begin{array}{lllllllllllllll}
\tilde{\rho}_{11}^{\prime(\pm 1)}&  \tilde{\rho}_{22}^{\prime(\pm 1)}& \tilde{\rho}_{00}^{\prime(\pm 1)}& \tilde{\rho}_{31}^{\prime(\pm 1)}& \tilde{\rho}_{13}^{\prime(\pm 1)}& \tilde{\rho}_{32}^{\prime(\pm 1)}& \tilde{\rho}_{23}^{\prime(\pm 1)}& \tilde{\rho}_{30}^{\prime(\pm 1)}& \tilde{\rho}_{03}^{\prime(\pm 1)}& \tilde{\rho}_{01}^{\prime(\pm 1)}& \tilde{\rho}_{10}^{\prime(\pm 1)}& \tilde{\rho}_{20}^{\prime(\pm 1)}& \tilde{\rho}_{02}^{\prime(\pm 1)}& \tilde{\rho}_{21}^{\prime(\pm 1)}& \tilde{\rho}_{12}^{\prime(\pm 1)}
\end{array}\right]^{T}\;,\label{eqa8}\\
B_{+}&=\left[\begin{array}{lllllllllllllll}
\frac{i}{\sqrt{2}}\tilde{\rho}_{13}^{(0)} & 0 & 0 & -\frac{i}{\sqrt{2}}\left(2\tilde{\rho}_{11}^{(0)}+\tilde{\rho}_{22}^{(0)}+\tilde{\rho}_{00}^{(0)}-1\right) & 0 & -\frac{i}{\sqrt{2}}\tilde{\rho}_{12}^{(0)} & 0 & -\frac{i}{\sqrt{2}}\tilde{\rho}_{10}^{(0)} & 0 & \frac{i}{\sqrt{2}}\tilde{\rho}_{03}^{(0)} & 0 & 0 & 0 & \frac{i}{\sqrt{2}}\tilde{\rho}_{23}^{(0)} & 0
\end{array}\right]^{T}\;,\label{eqa9}\\
B_{-} &=\left[\begin{array}{lllllllllllllll}
0& \frac{i}{\sqrt{2}}\tilde{\rho}_{23}^{(0)} & 0 & -\frac{i}{\sqrt{2}}\tilde{\rho}_{21}^{(0)} & 0 & -\frac{i}{\sqrt{2}}\left(\tilde{\rho}_{11}^{(0)}+2\tilde{\rho}_{22}^{(0)}+\tilde{\rho}_{00}^{(0)}-1\right) & 0 & -\frac{i}{\sqrt{2}}\tilde{\rho}_{20}^{(0)} & 0 & 0 & 0 & 0 & \frac{i}{\sqrt{2}}\tilde{\rho}_{03}^{(0)} & 0 & \frac{i}{\sqrt{2}}\tilde{\rho}_{13}^{(0)}
\end{array}\right]^{T}\;.\label{eqa10}
\end{align}
where $p_{11}^{\prime}=i\omega_{pc}-\gamma_{13}$, $p_{22}^{\prime}=i\omega_{pc}-\gamma_{23}$, $p_{00}^{\prime}=i\omega_{pc}-\gamma_{03}$, $p_{31}^{\prime}=i\left(\omega_{pc}+\Delta+\Delta_{z}\right)-\Gamma_{31}$, $p_{13}^{\prime}=-i\left(-\omega_{pc}+\Delta+\Delta_{z}\right)-\Gamma_{31}$, $p_{32}^{\prime}=i\left(\omega_{pc}+\Delta-\Delta_{z}\right)-\Gamma_{32}$, $p_{23}^{\prime}=-i\left(-\omega_{pc}+\Delta-\Delta_{z}\right)-\Gamma_{32}$, $p_{30}^{\prime}=i\left(\omega_{pc}+\Delta\right)-\Gamma_{30}$, $p_{03}^{\prime}=-i\left(-\omega_{pc}+\Delta\right)-\Gamma_{30}$, $p_{01}^{\prime}=i\left(\omega_{pc}+\Delta_{z}\right)-\Gamma_{01}$, $p_{10}^{\prime}=i\left(\omega_{pc}-\Delta_{z}\right)-\Gamma_{01}$, $p_{20}^{\prime}=i\left(\omega_{pc}-\Delta_{z}\right)-\Gamma_{20}$, $p_{02}^{\prime}=i\left(\omega_{pc}+\Delta_{z}\right)-\Gamma_{20}$, $p_{21}^{\prime}=i\left(\omega_{pc}-2\Delta_{z}\right)-\Gamma_{21}$, $p_{12}^{\prime}=i\left(\omega_{pc}+2\Delta_{z}\right)-\Gamma_{21}$.
\end{widetext}
The first three elements of the column vector $X_{0}$ provides the population of the atomic levels. The first order coherence terms $\tilde{\rho}_{31}^{\left(+1\right)}$ and $\tilde{\rho}_{32}^{\left(-1\right)}$ can be obtained from $X_{+}$ and $X_{-}$ respectively.

\end{document}